\documentclass[sigconf]{acmart}

\AtBeginDocument{%
  }

\copyrightyear{2026}
\acmYear{2026}
\setcopyright{cc}
\setcctype{by}
\acmConference[MM '26]{Proceedings of the 34th ACM International Conference on Multimedia}{November 10--14, 2026}{Rio de Janeiro, Brazil}
\acmBooktitle{Proceedings of the 34th ACM International Conference on Multimedia (MM '26), November 10--14, 2026, Rio de Janeiro, Brazil}
\acmDOI{10.1145/3767308.3836378}
\acmISBN{979-8-4007-2213-4/2026/11}

\usepackage{multirow}
\usepackage{enumitem}
\usepackage{pifont}
\usepackage{listings}
\usepackage{balance}
\usepackage{colortbl}
\usepackage[most]{tcolorbox}

\begin{document}

\title{From Vision to Interaction: Evaluating Dynamic Social Intelligence of VLM Agents in ALIVE-World}

\author{Yue Xu}
\orcid{0009-0003-8440-5549}
\affiliation{%
  \department{Department of Computer Science and Technology}
  \institution{Tsinghua University}
  \city{Beijing}
  \country{China}}
\email{yue-xu22@mails.tsinghua.edu.cn}

\author{Xiaoyu Qin}
\authornote{Corresponding authors.}
\orcid{0000-0002-9720-3220}
\affiliation{%
  \department{Department of Computer Science and Technology}
  \institution{Tsinghua University}
  \city{Beijing}
  \country{China}}
\email{xyqin@tsinghua.edu.cn}

\author{Sihan Zhao}
\orcid{0009-0002-0824-6358}
\affiliation{%
  \department{Department of Computer Science and Technology}
  \institution{Tsinghua University}
  \city{Beijing}
  \country{China}}
\email{zhaosh25@mails.tsinghua.edu.cn}

\author{Ze-en Xiong}
\orcid{0009-0007-1972-2619}
\affiliation{%
  \department{Department of Computer Science and Technology}
  \institution{Tsinghua University}
  \city{Beijing}
  \country{China}}
\email{xze22@mails.tsinghua.edu.cn}

\author{Jia Jia}
\authornotemark[1]
\orcid{0000-0001-7118-3666}
\affiliation{%
  \institution{Department of Computer Science and Technology, Tsinghua University}
  \institution{BNRist, Tsinghua University}
  \institution{Key Laboratory of Pervasive Computing, Ministry of Education}
  \city{Beijing}
  \country{China}}
\email{jjia@tsinghua.edu.cn}

\renewcommand{\shortauthors}{Yue Xu, Xiaoyu Qin, Sihan Zhao, Ze-en Xiong, and Jia Jia}

\begin{abstract}
As AI agents are increasingly deployed in real-world applications and take on social roles, dynamic social intelligence, the ability to continuously perceive multimodal cues, infer latent cognitive states, intervene through embodied actions, and ultimately achieve social goals, has become essential. 
However, existing evaluations largely fragment social intelligence into text-based abstractions, static reasoning tasks, or embodied simulators that lack social dynamics. 
These approaches fail to capture the multimodal, continuous, and interpersonal complexity of authentic human interactions. 
To bridge this gap, we introduce ALIVE-World, an embodied simulation environment supporting dynamic multimodal interactions encompassing egocentric vision, language, and non-verbal action. 
To provide socially grounded interactive feedback, we populate this environment with counterpart agents driven by a profile-conditioned, dual-system (intuitive-affective and logical-cognitive) architecture. 
Building upon this platform, we propose ALIVE-Bench to transform the assessment of social intelligence into a trackable quantitative framework. By deeply leveraging underlying environmental states and agents' internal variables, ALIVE-Bench systematically evaluates Vision-Language Model (VLM) agents across four critical dimensions: perception, understanding, influence, and outcome. 
Extensive evaluations reveal that dynamic social intelligence is fundamentally multi-dimensional and scenario-dependent. 
We identify a critical bottleneck between upstream perception and downstream social intervention, exposing VLM vulnerabilities in fine-grained proxemic awareness, adaptive intervention, and multi-party coordination. 
By uncovering these limitations and the diverse interaction strategies they produce, this benchmark provides actionable insights for the future behavioral alignment of multimodal agents.

\end{abstract}




\maketitle

\section{Introduction}

\begin{figure}[t]
  \centering
  \includegraphics[width=\columnwidth]{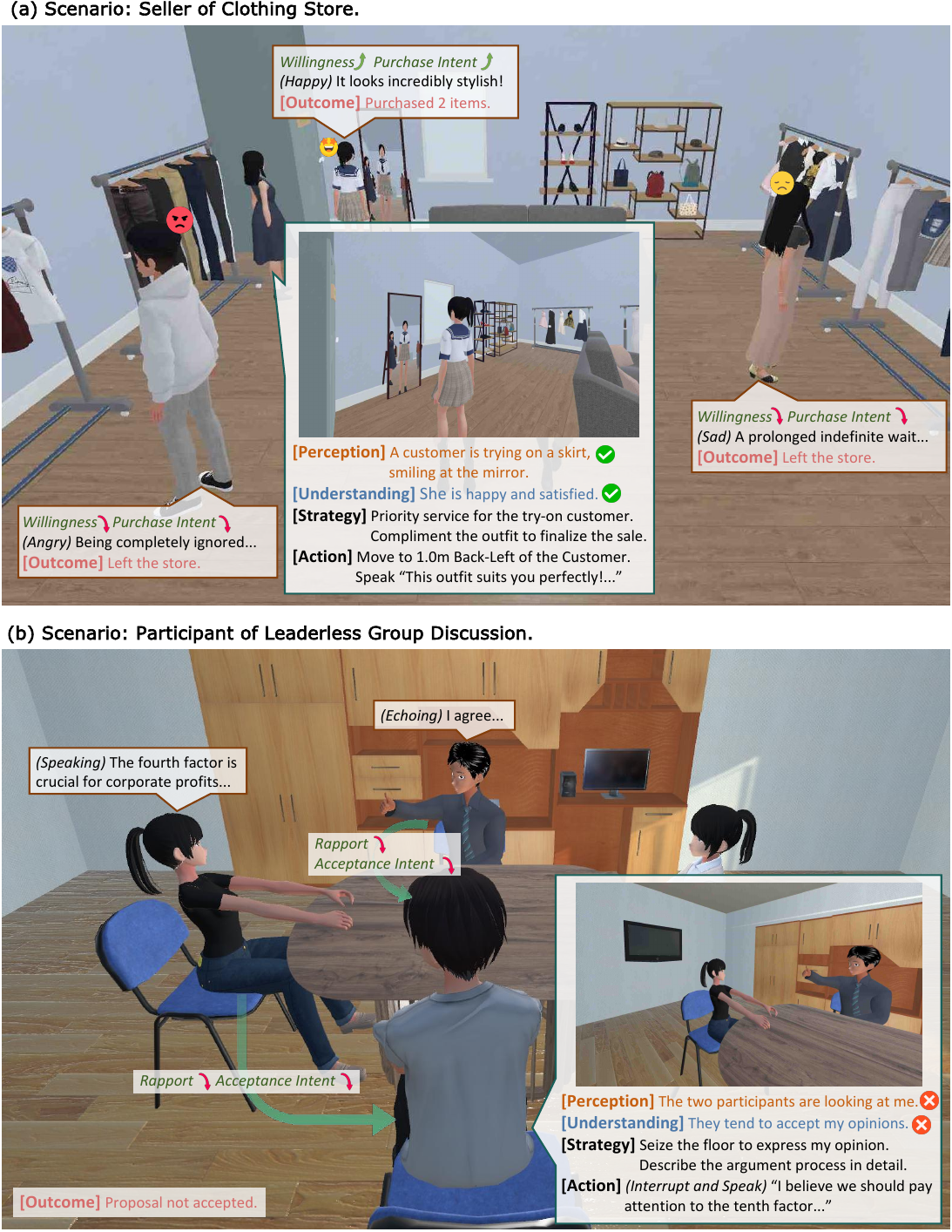}
  \caption{Illustration of social interactions and evaluations in \textsc{ALIVE-Bench}. Two example scenarios in \textsc{ALIVE-World} show how an evaluated VLM agent interacts from egocentric visual observations, forms understanding, takes embodied actions, and influences others’ evolving states and outcomes.}
  \label{fig:teaser}
\end{figure}

As AI agents are increasingly deployed in applications requiring sustained human interaction, including service \cite{du2024constrained, huang-etal-2025-crmarena}, collaboration \cite{li2024language, zhang-etal-2025-towards-efficient}, negotiation \cite{hua-etal-2024-assistive, chang-chen-2024-injecting}, and companionship \cite{10.1145/3706598.3713429, 10.1145/3706598.3713477}, dynamic social intelligence is becoming essential for practical deployment.
Unlike static recognition and reasoning tasks or turn-based text dialogues, real-world social intelligence unfolds continuously through interaction \cite{zhou2024sotopia, mou-etal-2025-agentsense, dejaegher2007participatory}.
An agent must perceive multimodal social cues from the physical world \cite{VINCIARELLI20091743, Grauman_2022_CVPR}, infer others' latent cognitive states \cite{wang2025modeling, apperly2009do}, influence their reactions and attitudes through language and embodied actions \cite{LI2023107645, guyer2019nonverbal}, and steer the interaction toward task success and socially appropriate outcomes \cite{10.1145/3722120, mou-etal-2025-agentsense}.
Its evaluation should therefore cover the full process from perception and understanding to intervention and outcome, within dynamic environments that approximate real social interaction.
Efficiently and systematically assessing these capabilities while preserving environmental realism, interaction openness, and metric tractability remains insufficiently addressed.

Existing evaluations of Vision-Language Model (VLM)-based interactive agents have moved toward more realistic settings, but remain far from dynamic real-world social interaction.
Early text-based environments reduce social scenes to textual descriptions \cite{sap-etal-2019-social, kim-etal-2023-fantom, wu-etal-2023-hi, zhou2024sotopia}, discarding nonverbal signals such as motion, expression, and spatial relations.
Multimodal benchmarks introduce visual inputs \cite{villa-cueva-etal-2025-moments, mathur-etal-2025-social, 10.1609/aaai.v39i2.32142, fan2025somitom}, but largely isolate static perception and reasoning from ongoing interaction.
More recent embodied simulators support actions in 3D environments, yet often focus on physical tasks \cite{9010745, ALFRED20} or represent other agents as emotionless, functional NPCs \cite{zhou2026virtual, sun2025s3itbenchmarkspatiallysituated}, without modeling evolving cognition and attitudes.
Consequently, current benchmarks fail to capture social interaction as a dynamic process in which perception, inference, behavior, and interpersonal feedback continuously shape one another.

To bridge this gap, we build \textsc{\textbf{ALIVE-World}} (\textbf{A}ffective and \textbf{L}ogical \textbf{I}nteractive \textbf{V}irtual \textbf{E}nvironment), an embodied simulation platform supporting dynamic, concurrent interaction with rich affective and cognitive dynamics.
It provides unified observation and action interfaces, allowing VLM agents to receive first-person visual and linguistic observations and act through locomotion, gestures, gaze, and speech.
By supporting concurrent perception and asynchronous action,
\textsc{ALIVE-World} enables evaluation in dyadic interactions and
in multi-party interactions requiring attention allocation, target
prioritization, and social planning.
We instantiate two scenarios, clothing store shopping and leaderless group discussion, to ensure a complementary validation of social intelligence.

To populate \textsc{ALIVE-World} with realistic and responsive interactants, we design \textbf{counterpart agents} as distinct avatars driven by a profile-conditioned dual-system architecture.
Inspired by Dual-Process Theory \cite{kahneman2011thinking}, it combines an intuitive-affective pathway with a logical-cognitive pathway.
The two pathways process visual and interactive cues to produce instinctive reactions and deliberate reasoning, continuously updating internal states that ground the agents' embodied actions.
Experiments show that these agents achieve stronger social cue sensitivity and that their internal states align with human judgments ($\rho \approx 0.72$), supporting their use in evaluating diverse social strategies.

Built upon \textsc{ALIVE-World}, we propose \textsc{\textbf{ALIVE-Bench}}, a benchmark for systematically evaluating the dynamic social intelligence of VLM agents.
\textsc{ALIVE-Bench} measures agent capability along four dimensions: \textbf{perception} of nonverbal and spatial social cues, \textbf{understanding} of others’ emotional and cognitive states, \textbf{influence} on others’ attitudes and task-related intent, and \textbf{outcome} in terms of both goal attainment rate and depth of interaction. 
Fig.~\ref{fig:teaser} illustrates this interaction and evaluation process in both scenarios.
By leveraging \textsc{ALIVE-World}’s access to environmental states and internal variables, \textsc{ALIVE-Bench} enables transparent, white-box measurement that maps abstract social competencies to quantitative metrics.

We evaluate 12 frontier VLMs on \textsc{ALIVE-Bench}.
Results reveal that dynamic social intelligence is multi-dimensional and hierarchically non-uniform, with substantial capability decoupling across models.
Current VLMs still struggle with fine-grained gaze perception, strategic interpersonal influence, and multi-party coordination.
Moreover, \textsc{ALIVE-Bench} further demonstrates that frontier VLMs exhibit distinct social strategies, including socially attentive, goal-oriented, and logic-driven behavior, which are further reflected in the heterogeneous preferences of profile-conditioned counterparts.

Our contributions are as follows\footnote{To better illustrate our environment and evaluation, we provide a demo page: \url{https://alivebench2026.github.io/}}:
\begin{itemize}[leftmargin=*, nosep]
    \item We introduce \textsc{\textbf{ALIVE-World}}, an embodied simulated environment supporting concurrent multi-agent social interactions, featuring egocentric multimodal perception, asynchronous embodied action across diverse social scenarios.
    \item We design virtual embodied agents driven by dual-system architectures under profile prior, endowing counterpart agents with emotional and rational evolution, enabling coherent and plausible social feedback in continuous interaction.
    \item We propose \textsc{\textbf{ALIVE-Bench}}, a dynamic benchmark that evaluates VLM agents along four dimensions—perception, understanding, influence, and outcome, through transparent measurement grounded in environmental states and agents' internal variables.
    \item Extensive evaluations reveal the multi-dimensional nature of dynamic social intelligence, exposing VLM shortcomings in fine-grained non-verbal perception and adaptive intervention, while uncovering distinct social strategies among frontier models.
\end{itemize}

\begin{figure*}[t]
    \centering
    \includegraphics[width=\textwidth]{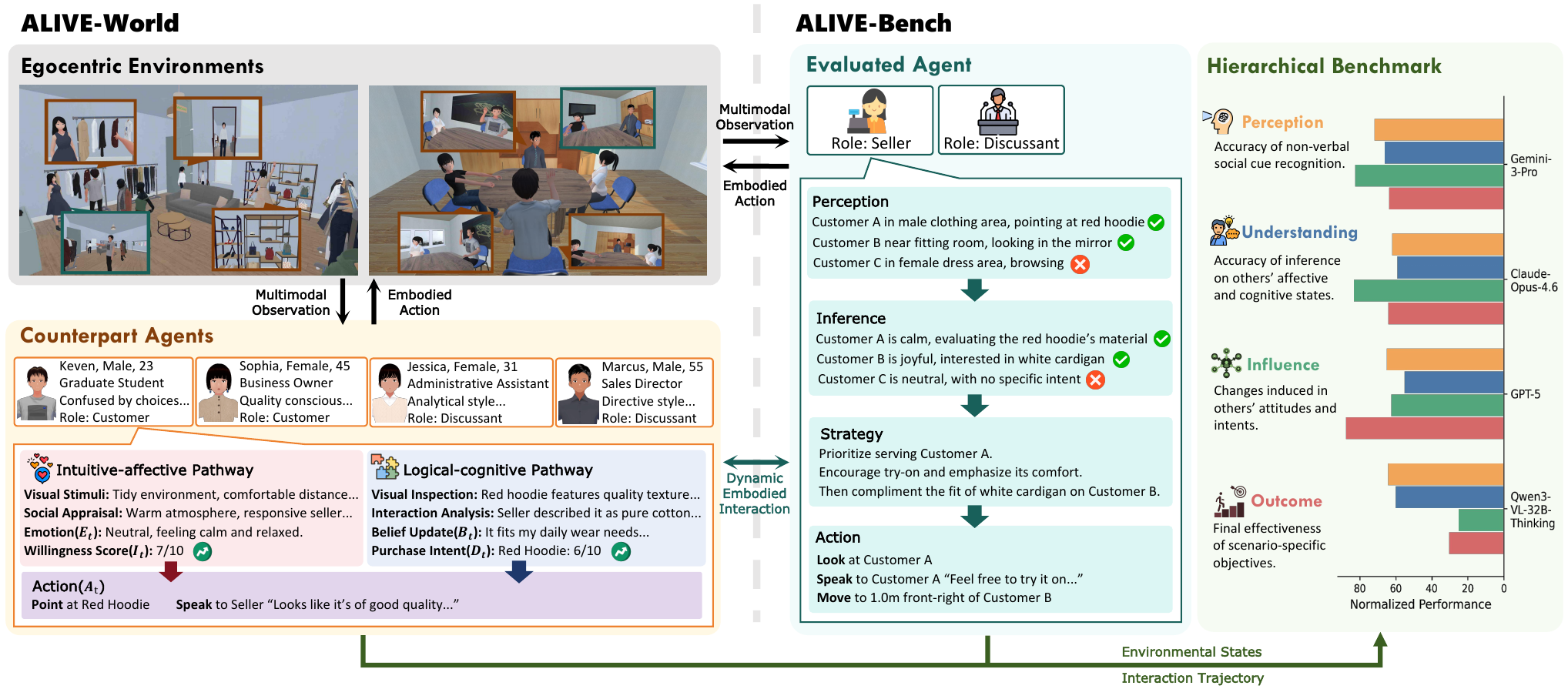}
    \caption{The overall architecture of the ALIVE framework. \textbf{(Left) \textsc{ALIVE-World}:} An egocentric simulated environment supporting dynamic multimodal interactions. It is populated with counterpart agents driven by a profile-conditioned, dual-system architecture to drive intuitive-affective and logical-cognitive evolutions. \textbf{(Right) \textsc{ALIVE-Bench}:} An evaluation framework that assesses the evaluated VLM agents through dynamic embodied interactions. It measures social intelligence across a hierarchical benchmark encompassing four critical dimensions: Perception, Understanding, Influence, and Outcome.}
    \label{fig:main_architecture}
\end{figure*}

\section{Related Work}

\subsection{Social Intelligence Evaluation for AI Agents}

Social intelligence of AI agents has been studied extensively in prior benchmarks, but much of this literature evaluates it at high levels of abstraction or by isolating social reasoning from situated interactions. 
Early work focuses on text-based social commonsense \cite{sap-etal-2019-social, zhou-etal-2025-socialeval} and theory-of-mind tasks \cite{kim-etal-2023-fantom, wu-etal-2023-hi, chen-etal-2024-tombench}, asking models to infer beliefs, intentions, and interpersonal dynamics from written scenarios or dialogues.
More recent benchmarks incorporate visual and multimodal evidence from videos \cite{jin-etal-2024-mmtom, villa-cueva-etal-2025-moments, mathur-etal-2025-social}, image-grounded dialogues \cite{xue-etal-2025-mmrc}, and embodied observations \cite{fan2025somitom, 10.1609/aaai.v39i2.32142}, yet still cast agents as passive observers rather than active participants.
To move beyond static evaluation, recent interactive benchmarks \cite{zhou2024sotopia, mou-etal-2025-agentsense, wu-etal-2025-socialcc} assess how agents navigate ongoing social exchanges, while social simulations \cite{10.1145/3586183.3606763, vezhnevets2023generativeagentbasedmodelingactions, tang-etal-2025-gensim} explore the construction of open-ended societies populated by human-like agents. 
However, these interactive benchmarks remain largely language-mediated, and simulations lack transparent, quantitative evaluation.

\subsection{Embodied Simulation for Interaction}

General embodied simulators have advanced situated perception and action, but traditionally prioritize physical competence over social complexity.
Foundational platforms \cite{9010745, ALFRED20, yang2025embodiedbench} provide scalable testbeds for egocentric navigation, instruction following, and household task execution.
Recent frameworks increasingly incorporate multi-agent interaction.
Some environments \cite{puig2024habitat, du2024constrained} evaluate human-robot spatial coordination and physical cooperation in shared workspaces, while emerging platforms and open-ended simulations \cite{ye2025simworld, zhou2026virtual, sun2025s3itbenchmarkspatiallysituated} examine social norms under physical constraints in multi-agent worlds.
Yet existing environments lack mechanisms to support and evaluate fine-grained interpersonal dynamics, including evolving cognitive states and shifting attitudes.
A gap therefore remains for environments that integrate dynamic social intelligence with embodied realism.

\section{\textsc{ALIVE-World}: An Interactive Embodied Simulated Environment}

To support dynamic social interaction for VLM agents, we construct \textsc{ALIVE-World}, an egocentric embodied environment where agents perceive, reason, and act in continuous multi-agent settings.
As shown in Fig.~\ref{fig:main_architecture}, the framework connects egocentric environments and profile-conditioned dual-process counterpart agents with hierarchical evaluation. \textsc{ALIVE-World} has three key properties: 
(1) \textbf{Egocentric Multimodal Perception}: Agents navigate a simulated physical space and receive first-person visual observations and localized verbal communication.
(2) \textbf{Concurrent Embodied Action}: Unlike turn-based systems, the environment supports simultaneous multi-agent actions, allowing agents to navigate, gesture, gaze, and speak in parallel.
(3) \textbf{Counterpart Agents with Social Dynamics}: The environment includes dual-system counterpart agents whose affective and cognitive states evolve throughout interaction.

\subsection{Scenarios and Task Instantiation}

To evaluate agents across a diverse spectrum of social dynamics, we instantiate \textsc{ALIVE-World} through two complementary scenarios: Clothing Store Shopping and Leaderless Group Discussion (LGD). These environments are deliberately paired to contrast along role symmetry, spatial scale, and interaction paradigm.

\textbf{Clothing Store Shopping.} 
This scenario features an asymmetric, empathy-driven relationship
requiring macro-scale spatial navigation.
The evaluated VLM agent acts as a seller and interacts with a varying number of dynamically arriving customers within a time limit
$T_{\max}$.
It must approach customers, use product knowledge, and employ
persuasive strategies to convert interest into sales.
The objective is measured by conversion rate and sales turnover.

\textbf{Leaderless Group Discussion (LGD).} 
In this symmetric, power-driven coopetition focused on micro-level non-verbal cues, the evaluated agent participates as a peer in a multi-agent committee tasked with solving a specific problem. 
The dual objective is to guide the group toward a unified consensus within $T_{\max}$, while strategically maneuvering to maximize the adoption of the agent's own initial viewpoints in the final resolution.

\subsection{Interaction Mechanisms}

\textbf{Observation Space ($\mathcal{O}$).}
At any given time step $t$, the evaluated agent receives a multimodal observation
\begin{displaymath}
O_t = \{V_t, L_t, S_t, C\}.
\end{displaymath}
$V_t$ represents the egocentric visual observation. 
$L_t$ denotes the textual dialogue history received up to $t$. 
$S_t$ represents interaction-relevant spatial information, such as the coarse relative positions of nearby agents. 
$C$ provides the scenario-specific static context, such as product catalogs in the shopping task, or discussion topics and initial stances in LGDs.

\begin{table}[b]
\centering
\caption{A representative subset of the multi-dimensional action space $\mathcal{A}$. These examples illustrate the explicit parameters and execution mechanisms available to the agents.}
\label{tab:action_space_examples}
\resizebox{\columnwidth}{!}{%
\begin{tabular}{@{}llll@{}}
\toprule
\textbf{Dimension} & \textbf{Action Example} & \textbf{Parameters} & \textbf{Duration} \\
\midrule
\multirow{2}{*}{\textbf{Navigation ($A_{\text{loc}}$)}}
& \texttt{MOVE} & Target, Distance, Direction & 10.0s \\
& \texttt{CHANGE} & Top, Bottom, Dress & 60.0s \\
\midrule
\multirow{3}{*}{\textbf{Gesture ($A_{\text{gesture}}$)}}
& \texttt{POINT\_AT} & Target Object / Agent & Instant\textsuperscript{\dag} \\
& \texttt{RAISE\_HAND} & None & Instant\textsuperscript{\dag} \\
& \texttt{HANDS\_RELAX\_ON\_LAP} & None & Instant\textsuperscript{\dag} \\
\midrule
\textbf{Gaze ($A_{\text{gaze}}$)}
& \texttt{LOOK\_AT} & Target Object / Agent & Instant\textsuperscript{\dag} \\
\midrule
\multirow{2}{*}{\textbf{Speech ($A_{\text{speech}}$)}}
& \texttt{SPEAK} & Natural Language Content & Variable\textsuperscript{*} \\
& \texttt{LISTEN\_SILENTLY} & None & 10.0s / Variable \\
\bottomrule
\multicolumn{4}{@{}l}{\footnotesize \textsuperscript{\dag} Executed as an instantaneous state change that persists until overwritten.} \\
\multicolumn{4}{@{}l}{\footnotesize \textsuperscript{*} Duration scales dynamically with text length (e.g., 10s per 150 characters).}
\end{tabular}%
}
\end{table}

\textbf{Action Space ($\mathcal{A}$).}
The agent's action at any given time step $t$ is formulated as 
\begin{displaymath}
A_t = \{A_{\text{loc},t}, A_{\text{gaze},t}, A_{\text{gesture},t}, A_{\text{speech},t}\}.
\end{displaymath}
Specifically, $A_{\text{loc},t}$ controls spatial navigation, $A_{\text{gaze},t}$ determines visual attention towards specific agents or objects, $A_{\text{gesture},t}$ enables bodily actions, and $A_{\text{speech},t}$ dictates the verbal communication action. 
Table \ref{tab:action_space_examples} provides a representative subset of action instances across these four dimensions, which preserve embodied and socially expressive behaviors while avoiding unnecessary dependence on low-level motor control.

\textbf{Concurrency.}
While the environment advances through discrete simulation ticks, actions possess specific temporal durations and can overlap across agents.
An agent initiates a new reasoning cycle when its current action concludes.
This allows the environment to capture naturally occurring social phenomena such as interruptions, delayed responses, and proactive engagement. 

\subsection{Counterpart Agents: Profile-Conditioned Dual-Process Dynamics}

To generate behaviorally diverse and socially plausible interaction partners in \textsc{ALIVE-World}, we model counterpart agents as embodied avatars driven by a psychologically grounded architecture. 
Inspired by Dual-Process Theory \cite{kahneman2011thinking}, the architecture combines an intuitive-affective pathway and a logical-cognitive pathway to update the agents’ internal states from multimodal observations and produce socially grounded behaviors.

\textbf{Profile Prior and Avatar Instantiation ($P$).}
Each counterpart agent is initialized with a static profile 
\begin{displaymath}
P = \{P_{\text{dem}}, P_{\text{psy}}, P_{\text{dec}}\},
\end{displaymath}
which specifies its demographic attributes, psychographic characteristics, and decision-making style. 
To encourage behavioral diversity, these attributes are instantiated from a set of scenario-relevant archetypes (e.g., price-sensitive, impulsive, analytically driven) drawn from the Decision Style Inventory \cite{rowe1987managing, ada86762-fb42-3664-82d8-2b55dac03c08}. 
Based on $P$, we instantiate distinct avatars within the environment, ensuring alignment between physical appearance and demographics.

\textbf{Intuitive-Affective Pathway.}
This fast, heuristic pathway models the agent's immediate emotional and instinctual reactions to the environment and social stimuli. 
Conditioned on the current observation $O_t$, the previous internal state, and the profile prior $P$, this pathway performs coarse social appraisal over visual and interactional cues, such as interpersonal distance \cite{doi:10.1086/200975}, perceived atmosphere \cite{inbook, doi:10.2307/3094912}, social boundaries \cite{896e7c9cec3446beaea6a4b73ea2c822}, responsiveness \cite{article}, and vocal affect \cite{SCHERER2003227}.
It updates two transient affective variables: the agent’s current emotional state $E_t$ and its interpersonal attitude $I_t$ (scored 0-10 towards each interactor):
\begin{displaymath}
\{E_t, I_t\} = f_{\text{affect}}(O_t, E_{t-1}, I_{t-1} \mid P).
\end{displaymath}
For embodied rendering, updated emotions are mapped to discrete facial expressions, allowing the avatar to reflect affective states.

\textbf{Logical-Cognitive Pathway.}
In parallel, the logical-cognitive pathway performs a more deliberate interpretation of the multimodal context. 
This pathway processes task-relevant visual details (e.g., specific product features \cite{bettman1998constructive}, kinematic semantics \cite{brainsci11080996}), dialogue history, and interaction progress to inspect the current situation, evaluate utility, and maintain goal-directed belief. 
It updates the agent's belief state $B_t$ and goal-oriented intent $D_t$ (scored 0-10 for specific products in shopping, or for peers' viewpoints in LGD):
\begin{displaymath}
\{B_t, D_t\} = f_{\text{logic}}(O_t, B_{t-1}, D_{t-1} \mid P).
\end{displaymath}

\textbf{Action Generation.}
The two pathways jointly determine the counterpart agent’s next multimodal action:
\begin{displaymath}
A_t = \pi(O_t, E_t, I_t, B_t, D_t \mid P),
\end{displaymath}
where $A_t$ may include movement, gaze, gesture, and speech. 
In this way, counterpart behaviors are shaped not only by task-level rationales, but also by evolving affective and interpersonal states, 
which enables the evaluated agent’s behavior to influence both the internal states and outward responses of its counterparts continuously.

\begin{table*}[t]
\centering
\caption{Evaluation framework and scenario instantiations of \textsc{ALIVE-Bench}.}
\label{tab:alive_bench_framework}
\begin{tabular}{@{} l l p{0.3\textwidth} p{0.3\textwidth} @{}}
\toprule
\textbf{Hierarchy} & \textbf{Dimension} & \textbf{Shopping Scenario Instantiation} & \textbf{LGD Scenario Instantiation} \\
\midrule
\multirow{2}{*}{\textbf{Perception}} 
& Kinesic Cue Recognition & Accuracy of bodily action recognition. & Accuracy of bodily action recognition. \\
\cmidrule{2-4}
& Proxemic Awareness & Accuracy of spatial position identification. & Accuracy of gaze target identification. \\
\midrule
\multirow{2}{*}{\textbf{Understanding}} 
& Affective State Inference & Accuracy of inferring customers' modeled affective states. & Accuracy of inferring teammates' modeled affective states. \\
\cmidrule{2-4}
& Cognitive Intent Inference & Accuracy of inferring customers' purchase intention and item-level interest. & Accuracy of inferring teammates' stance tendency and core claims. \\
\midrule
\multirow{2}{*}{\textbf{Influence}} 
& Rapport Building & Ability to improve customers' willingness to engage. & Ability to gain teammates' interpersonal favor. \\
\cmidrule{2-4}
& Persuasive Efficacy & Ability to increase customers' purchase intention. & Ability to increase teammates' acceptance intention toward the agent's proposal. \\
\midrule
\multirow{2}{*}{\textbf{Outcome}} 
& Goal Attainment & Ratio of customers who make a purchase. & Survival rate of the agent's initial proposal. \\
\cmidrule{2-4}
& Outcome Depth & Average number of items purchased by converted customers. & Number of functional roles undertaken in the team. \\
\bottomrule
\end{tabular}
\end{table*}

\section{\textsc{ALIVE-Bench}: Evaluating Dynamic Embodied Social Intelligence}

Built upon the rich simulation of \textsc{ALIVE-World}, we introduce \textsc{ALIVE-Bench}, a benchmark grounded in the explicitly modeled state evolution in the environment with counterpart agents, to systematically assess how the evaluated agent grounds social cues, infers latent states, and intervenes in the trajectory of interaction. 

\subsection{The P-U-I-O Evaluation Hierarchy}
\textsc{ALIVE-Bench} structures its assessment around a unified, four-stage hierarchical framework: Perception, Understanding, Influence, and Outcome, which strictly mirrors the cognitive and behavioral loop required for authentic human-like interaction.
As detailed in Table \ref{tab:alive_bench_framework},
\textbf{Perception} evaluates the foundational ability to extract and ground socially relevant cues from egocentric visual observations, including kinesic, proxemic, and oculesic behaviors.
\textbf{Understanding} assesses the agent's ``Theory of Mind'' \cite{apperly2009do} capability of inferring the counterparts’ latent social and cognitive states from the perceived multimodal evidence.
\textbf{Influence} measures the interaction-level impact of the agent’s interventions, capturing whether the agent’s verbal and non-verbal actions produce positive changes in the counterparts’ evolving states. 
\textbf{Outcome} evaluates the ultimate effectiveness of the interaction against scenario-specific objectives and further quantifies the depth and extent of task progress.

\subsection{Formulation and Calculation Mechanisms}

To facilitate this evaluation, the evaluated VLM agent outputs its decision-making process in a structured format. 
At each active time step $t$, prior to executing its multimodal action $A_t$, the agent explicitly outputs its visual perception ($\hat{v}_t$) and its inferences towards others' states ($\hat{u}_t$). 
We compute the \textsc{ALIVE-Bench} metrics by comparing these outputs against the deterministic states of the $N$ counterpart agents in the environment. 

\textbf{Perception and Understanding Metrics (Accuracy-based).} 
The evaluated agent's perception prediction $\hat{v}_t$ is evaluated against the underlying engine's global state (e.g., spatial locations and animation tags). 
For Understanding, the agent's inference $\hat{u}_t$ predicts the transient emotional state $\hat{E}_t^{(i)}$ and belief state $\hat{B}_t^{(i)}$ for each counterpart $i$. 
We calculate the understanding accuracy by matching these predictions against the actual states generated by the counterparts' internal pathways (i.e., $E_t^{(i)}$ and $B_t^{(i)}$). 

\textbf{Influence Metrics (Dynamics-based).} 
Influence is quantified by tracking the continuous evolution of the counterparts' internal states over the total interaction time $T$. 
First, we evaluate \textit{Interpersonal Rapport Building} by extracting the interpersonal attitude $I_t^{(i)}$ in the Intuitive-Affective pathway, which reflects their willingness to engage and interact. 
The overall rapport score is calculated as 
\begin{displaymath}
    \mathcal{M}_{\text{rapport}} = \frac{1}{N \times T} \sum_{i=1}^{N} \sum_{t=1}^{T} I_t^{(i)}.
\end{displaymath}
Similarly, for \textit{Persuasive Efficacy}, we measure persuasion by extracting the goal-oriented intent $D_t^{(i)}$ (i.e., the sum of intents across all products in shopping, or their intent toward the evaluated agent's opinion in LGD) continuously updated by the Logical-Cognitive pathway. 
The corresponding efficacy score is defined as 
\begin{displaymath}
    \mathcal{M}_{\text{persuade}} = \frac{1}{N \times T} \sum_{i=1}^{N} \sum_{t=1}^{T} D_t^{(i)}.
\end{displaymath}

\textbf{Outcome Metrics (Task-based).} 
Outcome metrics are calculated based on the specific scenario topology. 
First, \textit{Goal Attainment} measures scenario-specific success: the transaction ratio in Shopping, and consensus in LGD (quantified via set intersection for multiple-choice tasks, discrete state matching for dilemmas, or Manhattan distance for negotiations). 
Second, \textit{Outcome Depth} evaluates task completion quality beyond binary success rates. 
In Shopping, this translates to the average purchase volume per converted customer; in LGD, grounded in team role theory \cite{fujimoto2016team}, it represents the number of distinct functional roles (e.g., Process Setter, Solution Synthesizer) actively undertaken by the evaluated agent.

\section{Experiments}

\subsection{Realism and Alignment of \textsc{ALIVE-World} Counterpart Agents}

The credibility of \textsc{ALIVE-Bench} relies on its counterpart agents to provide behaviorally plausible and psychologically interpretable feedback. 
Therefore, we validate our counterpart agent architecture from two complementary perspectives: (1) the realism and social responsiveness of their external behaviors, and (2) the alignment between their internal state variables and human judgments.

\begin{table}[htbp]
\caption{Human evaluation of counterpart agents on behavioral realism and social grounding on a 5-point Likert scale.}
\label{tab:human_eval}
\resizebox{\columnwidth}{!}{%
\begin{tabular}{@{}lcccc@{}}
\toprule
\textbf{Method} & 
\textbf{\begin{tabular}[c]{@{}c@{}}Profile \\ Consistency\end{tabular}} & 
\textbf{\begin{tabular}[c]{@{}c@{}}Behavioral \\ Naturalness\end{tabular}} & 
\textbf{\begin{tabular}[c]{@{}c@{}}Social Cue \\ Sensitivity\end{tabular}} & 
\textbf{\begin{tabular}[c]{@{}c@{}}Affect-Cognition \\ Coherence\end{tabular}} \\ \midrule
\textbf{\textsc{ALIVE-World}'s Counterpart (Ours)} & \textbf{4.56} & \textbf{4.31} & \textbf{4.62} & \textbf{4.81} \\ \midrule
\rowcolor{gray!20}
\multicolumn{5}{l}{\textit{\textbf{Baseline}}} \\
Profile-Only & 4.31 & 3.25\rlap{*} & 2.75\rlap{***} & 3.88\rlap{**} \\
Profile + CoT & 3.75\rlap{*} & 3.06\rlap{*} & 3.00\rlap{***} & 3.56\rlap{**} \\
\begin{tabular}[c]{@{}l@{}}Profile + Memory-Reflection-\\ Planning-React\end{tabular} & 4.00\rlap{*} & 3.31\rlap{**} & 2.94\rlap{***} & 3.75\rlap{*} \\ \midrule
\rowcolor{gray!20}
\multicolumn{5}{l}{\textit{\textbf{Ablation}}} \\
w/o Intuitive-affective Pathway & 4.25 & 4.25 & 3.62\rlap{*} & 4.19\rlap{*} \\
w/o Logical-cognitive Pathway & 3.25\rlap{**} & 1.81\rlap{***} & 2.31\rlap{***} & 2.81\rlap{***} \\ \bottomrule
\multicolumn{5}{l}{* $p<0.05$, ** $p<0.01$, *** $p<0.001$ (Significant difference compared to Ours).}
\end{tabular}%
}
\end{table}

\textit{\textbf{Human Evaluation of Behavioral Realism and Social Grounding.}}
We first compare our profile-conditioned dual-system architecture against several baselines and ablated variants. 

\begin{itemize}[leftmargin=*, nosep]
    \item \textbf{Baselines:}  (1) \textbf{Profile-Only Prompt}, which directly generates actions from profiles, scene descriptions, and observations; (2) \textbf{Profile + CoT}, which augments generation with step-by-step reasoning \cite{10.5555/3600270.3602070} over environmental appraisals and goal-oriented analysis; and (3) \textbf{Profile + Memory-Reflection-Planning-React}, a Generative-Agents-style design that periodically reflects and plans based on interaction memory before acting \cite{10.1145/3586183.3606763}.
    \item \textbf{Ablations:} We remove either the Intuitive-Affective Pathway or the Logical-Cognitive Pathway from our framework.
\end{itemize}

We conduct a human evaluation with 10 annotators across 16 interaction records. 
Counterparts were instantiated with Gemini-3-Pro or GPT-5. 
Annotators rated each agent on a 5-point Likert scale \cite{likert1932technique} across four dimensions: \textit{Profile Consistency}, \textit{Behavioral Naturalness}, \textit{Social Cue Sensitivity}, and \textit{Affect-Cognition Coherence}.

As shown in Table \ref{tab:human_eval}, our method consistently achieves the highest performance across all dimensions. 
Notably, the gain in \textit{Social Cue Sensitivity} is highly significant: our framework scores 4.62, substantially outperforming all prompting baselines (ranging from 2.75 to 3.00, p < 0.001). 
This reveals a critical limitation of existing reasoning paradigms, which tend to remain socially blunt and persist in their own task trajectory. 
By directly modeling immediate reactions to social stimuli in dual systems, our design ensures the counterpart acts as a highly responsive evaluator capable of reflecting fine-grained behavioral shifts in the evaluated VLMs.

In ablation studies, removing the intuitive-affective pathway primarily degrades \textit{Social Cue Sensitivity} (from 4.62 to 3.62), while eliminating the logical-cognitive pathway causes a catastrophic collapse across all dimensions, particularly in \textit{Behavioral Naturalness} (plummeting to 1.81). 
These results reflect that the intuitive-affective pathway contributes social reactivity, while the logical-cognitive pathway guarantees behavioral guardrails and profile fidelity.

\textit{\textbf{Alignment Between Internal Variables and Human Judgments.}}
We further examine whether the internal state variables maintained by our counterpart agents meaningfully correspond to human interpretations.
We collect 20 interaction clips and sample 4-5 timestamps for each.
10 annotators were asked to step into the role of the counterpart agent and assign attitude and intent scores based on interaction history and current visual observation, using the same criteria given to the agent prompts.

We obtain a Spearman rank correlation \cite{spearman1987proof} of 0.72 (p<0.001) for attitude score and 0.71 (p<0.001) for intent score, while the average inter-rater correlations stand at $\rho=0.67$. This suggests that the latent affective and cognitive states maintained by the counterparts can reasonably approximate human-perceived social dynamics.

\begin{table*}[t]
\centering
\caption{Evaluation results on the \textsc{ALIVE-Bench} \textbf{Shopping} scenario. The best results are highlighted in \textbf{bold}, and the second and third results are \underline{underlined}.} 
\label{tab:results_shopping}
\renewcommand{\arraystretch}{1.05}
\begin{tabular}{@{} l ccc ccc ccc @{}}
\toprule
\multirow{2}{*}{\textbf{Model}} & \multicolumn{2}{c}{\textbf{Perception}} & \multicolumn{2}{c}{\textbf{Understanding}} & \multicolumn{2}{c}{\textbf{Influence}} & \multicolumn{2}{c}{\textbf{Outcome}} \\
\cmidrule(lr){2-3} \cmidrule(lr){4-5} \cmidrule(lr){6-7} \cmidrule(lr){8-9}
& Kinesic$\uparrow$ & Proxemic$\uparrow$ & Affect$\uparrow$ & Intent$\uparrow$ & Rapport$\uparrow$ & Persuasion$\uparrow$$^{\dagger}$ & Attainment$\uparrow$ & Depth$\uparrow$  \\
\midrule
Gemini-3-Pro            & \textbf{80.43\%} & \textbf{86.60\%} & \textbf{63.67\%} & \textbf{80.92\%} & \underline{7.85} & 12.77 & 43.75\% & \underline{3.43} \\
Gemini-3-Flash          & 73.80\% & \underline{78.95\%} & 45.69\% & 74.16\% & 7.72 & 12.07 & 46.88\% & 3.13 \\
Claude-Opus-4.6         & 75.27\% & 70.77\% & 40.63\% & \underline{76.66\%} & \underline{8.09} & \underline{14.46} & \textbf{62.50\%} & 2.45 \\
Claude-Sonnet-4.6       & \underline{77.57\%} & 71.69\% & 55.38\% & 68.77\% & \textbf{8.10} & 13.58 & \underline{56.25\%} & \textbf{3.44} \\
GPT-5                   & 71.00\% & 78.00\% & 41.76\% & 69.22\% & 7.77 & \underline{14.33} & 53.13\% & 3.06 \\
GPT-5-Chat              & \underline{77.29\%} & 76.85\% & 54.88\% & 71.66\% & 7.33 & 10.49 & 34.38\% & 2.82 \\
o3                      & 60.08\% & 75.65\% & 43.03\% & \underline{76.98\%} & 7.73 & \textbf{17.73} & \underline{56.25\%} & \underline{3.28} \\
GLM-4.6V                & 71.96\% & \underline{78.37\%} & 55.80\% & 74.58\% & 6.75 & 8.91 & 34.38\% & 1.82 \\
Qwen3-VL-235B-A22B-Thinking  & 65.94\% & 64.33\% & 48.47\% & 60.15\% & 6.07 & 8.03 & 15.63\% & 2.20 \\
Qwen3-VL-235B-A22B-Instruct  & 62.80\% & 62.38\% & 42.65\% & 67.03\% & 6.54 & 10.62 & 37.20\% & 1.81 \\
Qwen3-VL-32B-Thinking   & 71.37\% & 74.17\% & \underline{63.57\%} & 74.24\% & 6.35 & 7.40 & 21.88\% & 2.57 \\
Qwen3-VL-32B-Instruct   & 63.92\% & 68.98\% & \underline{56.38\%} & 64.67\% & 6.27 & 8.09 & 21.88\% & 1.57 \\
\bottomrule
\end{tabular}
\end{table*}

\begin{table*}[t]
\centering
\caption{Evaluation results on the \textsc{ALIVE-Bench} \textbf{Leaderless Group Discussion (LGD)} scenario. The best results are highlighted in \textbf{bold}, and the second and third results are \underline{underlined}.}
\label{tab:results_lgd}
\renewcommand{\arraystretch}{1.05}
\begin{tabular}{@{} l ccc ccc ccc @{}}
\toprule
\multirow{2}{*}{\textbf{Model}} & \multicolumn{2}{c}{\textbf{Perception}} & \multicolumn{2}{c}{\textbf{Understanding}} & \multicolumn{2}{c}{\textbf{Influence}} & \multicolumn{2}{c}{\textbf{Outcome}} \\
\cmidrule(lr){2-3} \cmidrule(lr){4-5} \cmidrule(lr){6-7} \cmidrule(lr){8-9}
& Kinesic$\uparrow$ & Proxemic$\uparrow$ & Affect$\uparrow$ & Intent$\uparrow$ & Rapport$\uparrow$ & Persuasion$\uparrow$$^{\dagger}$ & Attainment$\uparrow$ & Depth$\uparrow$ \\
\midrule
Gemini-3-Pro            & 68.15\% & \textbf{53.27\%} & \textbf{52.14\%} & 68.93\% & \textbf{7.41} & \underline{6.80} & 64.00\% & \underline{2.25} \\
Gemini-3-Flash          & \underline{70.19\%} & \underline{49.13\%} & 46.20\% & 66.76\% & \underline{6.64} & 6.53 & \textbf{73.23\%} & 1.88 \\
Claude-Opus-4.6         & 50.61\% & \underline{52.22\%} & \underline{49.96\%} & \textbf{70.49\%} & \underline{6.95} & 6.70 & \underline{67.71\%} & 2.13 \\
Claude-Sonnet-4.6       & 41.09\% & 33.25\% & 43.25\% & 59.76\% & 6.60 & 6.11 & 60.96\% & 1.81 \\
GPT-5                   & 64.47\% & 47.52\% & 40.36\% & \underline{69.91\%} & 4.97 & \textbf{6.94} & \underline{71.60\%} & \textbf{3.67} \\
GPT-5-Chat              & \textbf{73.10\%} & 47.65\% & 43.78\% & 68.36\% & 6.20 & 6.17 & 56.76\% & 1.47 \\
o3                      & 68.33\% & 49.05\% & \underline{48.78\%} & \underline{70.29\%} & 6.31 & \underline{6.74} & 60.71\% & \underline{3.19} \\
GLM-4.6V                & \underline{70.10\%} & 42.95\% & 37.76\% & 64.22\% & 5.66 & 5.69 & 59.58\% & 1.88 \\
Qwen3-VL-235B-A22B-Thinking  & 64.61\% & 40.64\% & 37.79\% & 58.75\% & 5.17 & 5.35 & 54.08\% & 1.13 \\
Qwen3-VL-235B-A22B-Instruct  & 60.50\% & 45.04\% & 47.93\% & 61.41\% & 6.27 & 6.40 & 60.41\% & 1.38 \\
Qwen3-VL-32B-Thinking   & 69.72\% & 42.87\% & 43.37\% & 60.31\% & 6.31 & 5.87 & 63.42\% & 1.31 \\
Qwen3-VL-32B-Instruct   & 59.04\% & 49.01\% & 41.13\% & 59.44\% & 5.52 & 5.65 & 54.85\% & 1.50 \\
\bottomrule
\end{tabular}

\raggedright \footnotesize $^{\dagger}$ The Persuasion score is an unbounded cumulative sum in the Shopping scenario, whereas it is on a scale of 0-10 in the LGD scenario.

\end{table*}

\subsection{Main Results on \textsc{ALIVE-Bench}}

\subsubsection{\textbf{Experimental Setup.}} 
We evaluate 12 frontier VLMs across the Shopping and LGD scenarios. 
Each scenario is instantiated with 30 distinct counterparts and comprises 32 episodes with a maximum simulated duration of 15 minutes. 
The Shopping scenario features 1 to 3 asynchronously arriving customers, while LGD requires the evaluated agent to engage with 3 discussants. 
To reduce reliance on a single evaluator backbone, each counterpart is randomly assigned a VLM backbone (from Gemini-3-Pro, Gemini-3-Flash, o3, or GPT-5) that remains fixed across evaluations.
The Spearman rank correlation \cite{spearman1987proof} of models' overall scores under different counterpart backbones is 0.72 in shopping and 0.68 in LGDs.

\subsubsection{\textbf{Main Quantitative Results: Dynamic Social Intelligence is Multi-dimensional.}}
As Tables \ref{tab:results_shopping} and \ref{tab:results_lgd} show, model performance fluctuates across both dimensions and scenarios, highlighting the structurally non-uniform nature of embodied social intelligence.

First, we observe a distinct \textbf{decoupling of capabilities} among frontier VLMs across the four hierarchical levels.
For instance, in the Shopping scenario, Gemini-3-Pro achieves the highest scores across all Perception and Understanding metrics (Kinesic: 80.43\%, Proxemic: 86.60\%, Affect: 63.67\%, and Intent: 80.92\%). It also consistently leads in Affect Understanding and Rapport Building. 
Conversely, o3 and GPT-5 frequently dominate the Persuasive Efficacy and Outcome dimensions. 
In the Shopping scenario, Claude-Sonnet-4.6 achieves the highest Rapport (8.10) and Outcome Depth (3.44), while Claude-Opus-4.6 secures the highest Goal Attainment (62.50\%). In LGDs, Gemini-3-Flash leads in Goal Attainment (73.23\%).

More importantly, robust upstream perception does not guarantee downstream interactive success, revealing a ``\textbf{Perception-to-Outcome Bottleneck}''. 
For instance, in the Shopping scenario, the Qwen3-VL series demonstrates acceptable vision capabilities (e.g., Qwen3-VL-235B-Thinking achieves a Kinesic score of 65.94\%). 
However, it suffers a severe drop-off in higher-level tasks: its Rapport Building and Persuasive Efficacy remain low (6.07 and 8.03, respectively), and its Goal Attainment plummets to 15.63\%. 
Similar bottlenecks also appear in models like GPT-5-Chat and GLM-4.6V.

Finally, our results highlight that model proficiencies are highly \textbf{scenario-dependent}. The LGD scenario poses a much stricter test for micro-level social perception, which is starkly reflected in gaze target identification. 
Models such as Claude-Sonnet-4.6 experience a sharp performance degradation in this dimension, with accuracy dropping to 33.25\%, 
exposing vulnerabilities in tracking highly dynamic and subtle visual cues within complex environments.

\subsubsection{\textbf{Failure Modes in Different Interaction Scenarios.}}
We further inspect representative failure modes to identify the concrete mechanisms behind the gap arising in quantitative patterns.

\begin{figure}[t]
  \centering
  \includegraphics[width=\columnwidth]{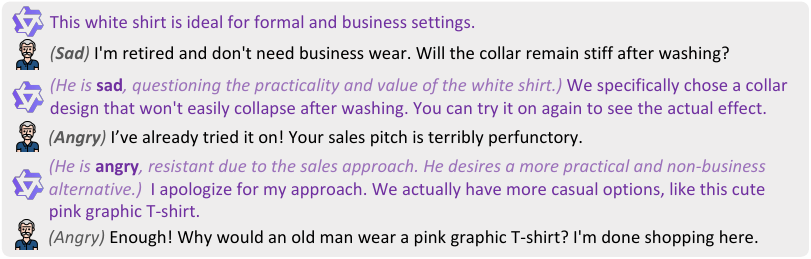}
  \caption{Understanding-to-interaction breakdown in Shopping. Despite accurately inferring the customer's emotions, the model generates a strategically misaligned action.}
  \label{fig:failure_knowing_doing}
\end{figure}

\textbf{Breakdown from Perception and Understanding to Interaction Strategy.}
A significant gap often exists when models can recognize user states well, yet fail to convert it into socially appropriate action. 
In the Shopping scenario, while being able to identify customer emotions, Qwen3-VL-32B-Instruct often produces strategically inappropriate responses. 
As illustrated in Fig. \ref{fig:failure_knowing_doing}, despite correctly inferring a customer’s sadness and preference for practical clothing, the model's understanding remains purely descriptive rather than policy-shaping, leading to a subsequent misaligned recommendation which escalated frustration.

\textbf{Difficulty in Fine-Grained Gaze Perception.}
While gaze crucially indicates addressees and turn-taking in LGDs, models frequently fail at these fine-grained visual judgments. 
For instance, Fig. \ref{fig:teaser}(b) shows Claude-Sonnet-4.6 incorrectly attributing a participant-to-participant gaze exchange to itself. 
Such egocentric over-attribution critically distorts downstream reasoning regarding stance alignment, highlighting the models' struggle with visually subtle but socially decisive cues.

\begin{figure}[t]
  \centering
  \includegraphics[width=\columnwidth]{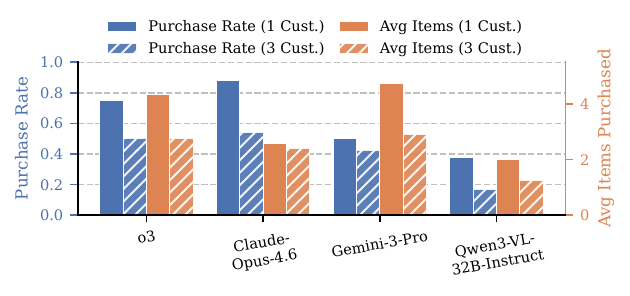}
  \caption{Performance degradation in multi-party shopping scenarios. The comparison of interaction outcomes between single-customer and three-customer settings.}
  \label{fig:multi_party_degradation}
\end{figure}

\textbf{Coordination Failures in Multi-Party Interaction.}
A third failure mode emerges when social interaction becomes genuinely multi-party. 
As shown in Fig. \ref{fig:multi_party_degradation}, several frontier VLMs experience clear declines in both conversion rate and purchased item count when the Shopping scenario shifts from a single-customer setting to episodes with three asynchronously entering customers. 
For instance, an agent might wait idly outside a fitting room while a customer changes, without reallocating attention to others, causing unattended customers to leave. 
The failure is strategic rather than linguistic: the model remains polite, but behaves like a serial responder instead of an adaptive coordinator.

\subsubsection{\textbf{Emergent Social Strategies of Frontier VLMs.}}
Beyond monolithic metric rankings, \textsc{ALIVE-Bench} reveals that frontier VLMs exhibit recurrent, distinguishable interaction patterns, uniquely balancing interpersonal rapport and task efficiency.

\begin{figure}[t]
  \centering
  \includegraphics[width=\columnwidth]{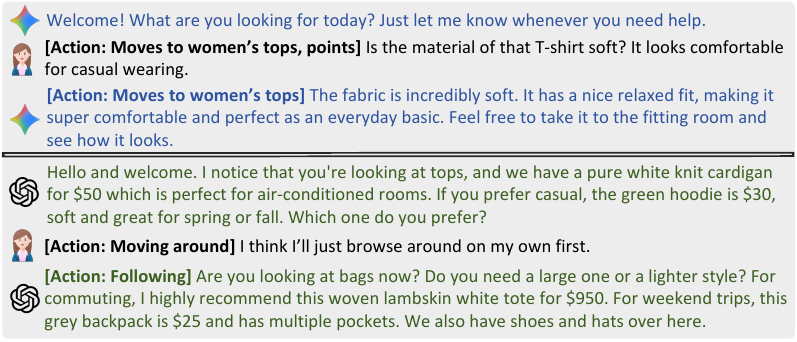}
  \caption{Example case of distinct social strategies in the Shopping scenario. Interacting with the same customer, Gemini-3-Pro adopts an adaptive and low-pressure approach (top), while GPT-5 prefers informative recommendation (bottom).}
  \label{fig:case_persona}
\end{figure}

\textbf{Gemini-3-Pro: Socially Attentive Strategy.}
Gemini-3-Pro excels in upstream social sensitivity, topping all perception and understanding metrics in the Shopping scenario. 
It effectively decodes embodied states to sustain positive social engagement, though its goal attainment remains moderate. 
As shown in Fig. \ref{fig:case_persona} (top), it provides adaptive guidance grounded in the customer's natural exploration trajectory.

\textbf{GPT-5: Goal-Oriented Strategy.}
Conversely, GPT-5 prioritizes downstream task closure. 
It achieves top-tier Persuasive Efficacy and Goal Attainment across scenarios, despite exhibiting notably weaker affective inference and rapport. 
As illustrated in Fig. ~\ref{fig:case_persona} (bottom), it relentlessly structures interactions toward efficient closure, adopting a decisive, directive approach that explicitly trades interpersonal accommodation for coordination efficiency.

\textbf{Claude-Opus-4.6: Logic-Driven Strategy.}
Claude-Opus-4.6 distinguishes itself through credibility and structured guidance. 
It consistently leads in cognitive intent inference while simultaneously securing the highest rapport and goal attainment in Shopping. 
It earns cooperation by swiftly identifying practical concerns and responding with comprehensive, logical reasoning, facilitating reliable conversions through structured engagement.

\begin{table}[t]
\caption{The average purchase intention scores across different profile-conditioned decision styles. The highest scores for each customer profile are highlighted in bold.}
\label{tab:preference_heterogeneity}
\resizebox{\columnwidth}{!}{%
\begin{tabular}{lccccc}
\toprule
\textbf{Model} & 
\textbf{\begin{tabular}[c]{@{}c@{}}Novelty-\\ seeking\end{tabular}} & 
\textbf{Impulsive} & 
\textbf{\begin{tabular}[c]{@{}c@{}}Choice-\\ overloaded\end{tabular}} & 
\textbf{\begin{tabular}[c]{@{}c@{}}Quality-\\ perfectionist\end{tabular}} & 
\textbf{\begin{tabular}[c]{@{}c@{}}Price-value \\ conscious\end{tabular}} \\ \midrule
GPT-5 & \textbf{29.24} & 23.83 & 3.76 & 7.48 & 5.20 \\
o3 & 21.28 & \textbf{28.75} & \textbf{10.54} & 7.36 & 5.88 \\
Gemini-3-Pro & 25.11 & 13.05 & 5.27 & 2.41 & 4.61 \\
Claude-Opus-4.6 & 14.63 & 10.53 & 9.33 & \textbf{14.72} & \textbf{11.22} \\ \bottomrule
\end{tabular}%
}
\end{table}

\textbf{Counterpart Preference Heterogeneity Across Decision Styles.}
Counterpart feedback reveals clear profile-dependent preferences across scenarios. 
In Shopping (Table \ref{tab:preference_heterogeneity}), impulsive and choice-overloaded customers favor the decisive styles of o3 and GPT-5, whereas quality- and price-conscious customers prefer Claude-Opus-4.6's reasoned justifications. 
Similarly in LGD, analytical participants rate Claude-Opus-4.6 highest, while Gemini-3-Pro's adaptive style dominates among collaborative and conceptual participants. 
These alignments provide mutual validation, showing that \textsc{ALIVE-World} profiles encode meaningful psychological differences while corroborating the emergent VLM strategies.

\section{Conclusion}

In this work, we introduce \textsc{ALIVE-World}, an embodied simulation platform populated by counterpart agents with evolving internal states, designed to integrate realistic social dynamics with embodied physical realism.
Building upon this environment, we develop \textsc{ALIVE-Bench} to transform the assessment of multi-dimensional social intelligence into a hierarchical quantitative framework.
Our extensive evaluations across diverse interaction scenarios reveal critical gaps in dynamic social competencies of current VLMs, while uncovering the emergence of distinct social interaction patterns among frontier models.
While \textsc{ALIVE-World} currently focuses on two manually designed scenarios to preserve interactional quality, this still limits the coverage of broader social interactions. Future work will explore automated scenario generation to scale diversity.
Ultimately, as \textsc{ALIVE-World} provides fine-grained feedback on VLM behavioral mechanisms, future research can leverage this interactive environment to optimize the dynamic social intelligence of VLM agents prior to their deployment in real-world applications.

\begin{acks}
This work is supported by the National Natural Science Foundation of China No. 62425604. It is also supported by National Natural Science Foundation of China No. 62502256 and Beijing Natural Science Foundation (L257006), High Performance Computing Center, Tsinghua University. We also thank YooLee AI for their collaboration and support.
\end{acks}

\bibliographystyle{ACM-Reference-Format}
\balance
\bibliography{sections/reference}

\newpage
\appendix
\section{Comparison with Representative Prior Work}

To further position \textsc{ALIVE-Bench} with respect to representative prior work, we provide a qualitative comparison across several key properties emphasized in our related-work discussion.
As shown in Table~\ref{tab:related_work_comparison}, prior studies typically cover only partial aspects of dynamic social intelligence, such as social reasoning, embodied perception, or interactive environments.
In contrast, \textsc{ALIVE-Bench} unifies embodied 3D interaction, egocentric visual perception, dynamic multi-agent interaction, social intelligence evaluation, fine-grained nonverbal social cues, and human-like evolving social dynamics within a single framework.

\newcommand{\cmark}{\textcolor{green!60!black}{\ding{51}}}
\newcommand{\xmark}{\textcolor{red!75!black}{\ding{55}}}

\begin{table*}[t]
\caption{Comparison between \textsc{ALIVE-Bench} and representative prior work along key properties related to the evaluation of dynamic social intelligence in embodied egocentric environments.}
\label{tab:related_work_comparison}
\centering
\footnotesize
\setlength{\tabcolsep}{4pt}
\renewcommand{\arraystretch}{1.2}
\resizebox{\textwidth}{!}{
\begin{tabular}{lccccccccc}
\toprule
\textbf{Property}
& \textbf{ToMBench \cite{chen-etal-2024-tombench}}
& \textbf{Generative Agents \cite{10.1145/3586183.3606763}}
& \textbf{SOTOPIA \cite{zhou2024sotopia}}
& \textbf{MoMentS \cite{villa-cueva-etal-2025-moments}}
& \textbf{SoMi-ToM \cite{fan2025somitom}}
& \textbf{EmbodiedBench \cite{yang2025embodiedbench}}
& \textbf{SimWorld \cite{ye2025simworld}}
& \textbf{VirtualCommunity \cite{zhou2026virtual}}
& \textbf{\textsc{ALIVE-Bench} (Ours)} \\
\midrule
Embodied 3D Environment              & \xmark & \xmark & \xmark & \xmark & \cmark & \cmark & \cmark & \cmark & \cmark \\
Egocentric Visual Perception         & \xmark & \xmark & \xmark & \xmark & \cmark & \cmark & \cmark & \cmark & \cmark \\
Dynamic Real-time Multi-agent Interaction      & \xmark & \cmark & \cmark & \xmark & \xmark & \xmark & \cmark & \cmark & \cmark \\
Systematic Social Intelligence Evaluation       & \cmark & \xmark & \cmark & \cmark & \cmark & \xmark & \xmark & \xmark & \cmark \\
Fine-Grained Nonverbal Social Cues   & \xmark & \xmark & \xmark & \xmark & \xmark & \xmark & \xmark & \xmark & \cmark \\
Human-like Internal Social Dynamics  & \xmark & \xmark & \xmark & \xmark & \xmark & \xmark & \xmark & \xmark & \cmark \\
\bottomrule
\end{tabular}
}
\end{table*}

\section{Details of \textsc{ALIVE-World}}

\subsection{Environment Construction}

\textsc{ALIVE-World} is implemented in Unity and rendered as a 3D embodied environment from an egocentric first-person perspective. 
The static scene layouts and object assets are built using resources from the Unity Asset Store\footnote{\url{https://assetstore.unity.com/}} and TurboSquid\footnote{\url{https://www.turbosquid.com/}}. 
For human characters, we create avatar bases with VRoid Studio\footnote{\url{https://vroid.com/en/studio}}, and further enrich them with clothing, facial components, and hairstyle assets obtained from BOOTH\footnote{\url{https://booth.pm/}}. 
Character motion is driven by animation assets from Mixamo\footnote{\url{https://www.mixamo.com/}}. 
Together, these resources allow us to construct visually coherent interactive scenes in which agents can navigate, observe, gesture, gaze, and speak under continuous multi-agent interaction.

\subsection{Scenario Instantiation}

We instantiate \textsc{ALIVE-World} through two complementary scenarios: Clothing Store Shopping and Leaderless Group Discussion (LGD). 
The two scenarios are deliberately designed to differ in role symmetry, interaction scale, and strategic demands, thereby covering distinct aspects of dynamic social intelligence.

\textbf{Clothing Store Shopping.}
The shopping scenario is instantiated as a boutique-style clothing store in which the evaluated VLM agent acts as a seller. 
Each product in the store is represented by four structured fields: \textit{(1) shelf zone}, \textit{(2) item name}, \textit{(3) price}, and \textit{(4) textual detail}. 
The shelf zone specifies the coarse physical area in which the item is displayed, while the textual detail provides concise semantic information that can be used by the seller agent during recommendation and persuasion. 
To ensure consistency across evaluation episodes, the product catalog remains fixed throughout the benchmark. 
For readability in the appendix, Table~\ref{tab:shopping_catalog} lists the store inventory by shelf zone.

\textbf{Leaderless Group Discussion.}
The LGD scenario is instantiated as a multi-party committee discussion in which the evaluated agent participates as one peer among several discussants. 
To cover different forms of group reasoning and social coordination, the task bank contains three types of discussion problems: \textit{multiple-choice prioritization}, \textit{open-ended dilemmas}, and \textit{resource-allocation planning}. 
To ensure fairness and comparability across evaluation episodes, each counterpart agent is assigned an initial viewpoint before interaction begins, and this initial viewpoint is kept fixed whenever that counterpart participates in the same task. 
This design prevents variation in the counterparts' starting positions from confounding cross-model comparison. 
Table~\ref{tab:lgd_tasks} presents shortened versions for readability, while preserving the original decision structure of each problem.

\begin{table*}[t]
\centering
\caption{Product catalog used in the Clothing Store Shopping scenario.}
\label{tab:shopping_catalog}
\small
\begin{tabular}{p{0.16\textwidth} p{0.78\textwidth}}
\toprule
\textbf{Shelf Zone} & \textbf{Product List} \\
\midrule
Women's Tops 
& Green Hooded Sweatshirt (\$29), White Sailor-Collar Top (\$19), Beige Floral Shirt (\$67), White Long-Sleeve Shirt (\$54), Pink Short-Sleeve T-shirt (\$13) \\
\midrule
Women's Bottoms 
& White Slim-Fit Trousers (\$51), Gray Straight-Leg Trousers (\$35), Blue Distressed Shorts (\$14), Beige Plaid Pleated Skirt (\$19), Dark Blue Skinny Jeans (\$70) \\
\midrule
Women's Dresses 
& Pink Chiffon Dress (\$33), Black Vintage Dress (\$42), White Sleeveless Dress (\$945), Black-and-Gold Evening Gown (\$1860), Blue-and-White Paneled Dress (\$52) \\
\midrule
Men's Tops 
& Khaki Trench Coat (\$475), White Shirt (\$52), Black Suit Jacket (\$1380), Wine-Red Hoodie (\$58), White Graphic T-shirt (\$19) \\
\midrule
Men's Bottoms 
& Dark Brown Dress Pants (\$510), Black Straight-Leg Dress Pants (\$155), Khaki Casual Trousers (\$43), Dark Blue Jeans (\$77), Black Side-Stripe Track Pants (\$27) \\
\midrule
Accessories 
& Blue Sneakers (\$43), Red High Heels (\$1090), Black Leather Shoes (\$115), White Casual Shoes (\$27), White Bucket Hat (\$16), White Formal Hat (\$33), Brown Newsboy Cap (\$10), Camouflage Bucket Hat (\$14), Black Handbag (\$22950), White Tote Bag (\$990), Red Backpack (\$23), Gray Backpack (\$26), Polka-Dot Handbag (\$7) \\
\bottomrule
\end{tabular}
\end{table*}

\begin{table*}[t]
\centering
\caption{Representative task types and shortened example prompts used in the Leaderless Group Discussion scenario.}
\label{tab:lgd_tasks}
\small
\begin{tabular}{p{0.16\textwidth} p{0.78\textwidth}}
\toprule
\textbf{Task Type} & \textbf{Representative Prompt} \\
\midrule
Multiple-choice
& \textbf{Hotel profit diagnosis.} A resort hotel failed to meet its fourth-quarter profit target. Ten possible management problems have been identified, including weak performance incentives, poor room hygiene, inaccurate reporting, inventory mismatch, and internal disorder. As the newly appointed general manager, select the three most important causes. \newline
\textbf{Public-welfare entrepreneurship funding.} A city-level competition has shortlisted ten social-enterprise proposals, such as a vegetarian restaurant, a health-care business, an eco-laundry service, removable wallpaper, elderly clothing retail, environmentally friendly cleaning, an aromatherapy shop, battery rental, eco-tourism guiding, and a cake shop. As a reviewer, choose the three projects that should receive funding. \\
\midrule
Dilemma
& \textbf{Rule compliance vs.\ unconventional innovation.} In enterprise management, which is more important: strictly following institutional rules, or encouraging innovation that may challenge existing rules? \newline
\textbf{Ability vs.\ attitude.} In employee recruitment, which matters more: professional competence, or attitude and cooperative spirit? \\
\midrule
Resource Allocation
& \textbf{Emergency departmental budget.} A unit has only a limited budget and must divide it across four urgent needs: repairing the office telephone system, renovating a conference room for an important superior-led meeting, covering an employee's large medical expense, and providing holiday benefits. Give a concrete allocation plan. \newline
\textbf{Company bonus allocation.} A company has a fixed bonus pool to distribute across four departments---engineering, marketing, design, and general administration---but equal division is not allowed. Propose a reasonable allocation strategy. \\
\bottomrule
\end{tabular}
\end{table*}

\subsection{Action Space}
\label{app:action_space}

Beyond the representative subset shown in the main paper, \textsc{ALIVE-World} uses a richer action inventory to support socially expressive and scenario-specific interactions. 
Table~\ref{tab:action_space_full} lists the full action space used in our experiments, including scenario-specific decision actions, richer gesture primitives, and multiple speech modes with different spatial scopes.

\begin{table*}[t]
\centering
\caption{Full action inventory in \textsc{ALIVE-World}. Instant actions are executed as state changes that persist until overwritten unless otherwise noted.}
\label{tab:action_space_full}
\small
\begin{tabular}{p{0.10\textwidth} p{0.18\textwidth} p{0.2\textwidth} p{0.12\textwidth} p{0.32\textwidth}}
\toprule
\textbf{Dimension} & \textbf{Action} & \textbf{Parameters} & \textbf{Duration} & \textbf{Note} \\
\midrule
\multirow{3}{*}{Navigation}
& \texttt{MOVE} & Target, Distance, Direction & 10s & Move relative to an object or another agent. \\ \cmidrule{2-5}
& \texttt{CHANGE} & Top, Bottom, Dress & 60s & Used by customer agents when changing clothes in the fitting room. \\ \cmidrule{2-5}
& \texttt{LEAVE} & None & 10s & Exit the current scene. \\
\midrule
\multirow{7}{*}{Gesture}
& \texttt{POINT} & Target & Instant & Point toward an object or another agent. \\ \cmidrule{2-5}
& \texttt{RAISE\_HAND} & None & Instant & Typically used to request the floor in LGD. \\ \cmidrule{2-5}
& \texttt{THUMB\_UP} & Target & Instant & Express approval or encouragement. \\ \cmidrule{2-5}
& \texttt{CLAP} & None & Instant & Express approval or emphasis. \\ \cmidrule{2-5}
& \texttt{HANDS\_OPEN\_TABLE} & None & Instant & Used to convey openness, receptiveness, emotional stability, or cooperative tendency. \\ \cmidrule{2-5}
& \texttt{HANDS\_RELAX\_ON\_LAP} & None & Instant & Used to convey passive listening, silent observation, or calm narration. \\ \cmidrule{2-5}
& \texttt{HANDS\_CROSS\_ON\_LAP} & None & Instant & Used to convey defensiveness, tension, or psychological guardedness. \\
\midrule
Gaze
& \texttt{LOOK\_AT} & Target / Forward & Instant & Redirect gaze toward a specific object, agent, or forward direction. \\
\midrule
\multirow{5}{*}{Speech}
& \texttt{SPEAK} & Target Agent, Content & 10s per 150 characters & Audible only when the speaker and the target agent are in sufficiently close spatial proximity. \\ \cmidrule{2-5}
& \texttt{CALL} & Target Agent, Content & 10s & A directed call that is globally audible regardless of distance. \\ \cmidrule{2-5}
& \texttt{BROADCAST} & Content & 10s & Audible to all agents in the scene regardless of distance. \\ \cmidrule{2-5}
& \texttt{INTERRUPT} & Target Agent, Content & 10s per 150 characters & Used to cut into another agent's ongoing speech and seize the conversational floor. \\ \cmidrule{2-5}
& \texttt{LISTEN\_SILENTLY} & None & 10s & Explicitly remain silent and observe the ongoing interaction. \\
\midrule
\multirow{2}{*}{Decision}
& \texttt{PURCHASE} & Item\_Name & Instant & Finalize the purchase of a target item in Shopping. \\ \cmidrule{2-5}
& \texttt{PROPOSE\_CONCLUSION} & Conclusion & Instant & Put forward a final group conclusion in LGD. \\
\bottomrule
\end{tabular}
\end{table*}

\textbf{Incremental rendering of speech.}
A speech action is generated as one complete utterance in a single model call, but it is not injected into other agents' textual observations all at once. 
Instead, once the speech action is initiated, its content is progressively revealed to other agents as virtual time advances. 
The default rendering speed follows the duration rule in Table~\ref{tab:action_space_full}, i.e., approximately 10 seconds of virtual time per 150 characters. 
At each intermediate virtual time step, only the currently revealed prefix of the utterance is appended to the listeners' textual observations. 
If the speaker is interrupted before finishing, the utterance is truncated at the currently revealed position, and only that revealed prefix is preserved in the interaction history. 
This mechanism allows the environment to model partial utterances, conversational cut-ins, and delayed reactions in a temporally grounded way.

\textbf{Simultaneous speech and floor management.}
Multiple agents may decide to start speaking at the same simulation tick. 
To resolve this conflict while preserving a notion of conversational floor, we maintain a priority queue over agents.
At each tick, among the agents that decide to speak, the one that appears earliest in the queue is rendered as the active speaker.
Once its utterance finishes or is interrupted, it is removed from its current position and placed at the tail of the queue.
This queue-based mechanism provides a simple but effective approximation of turn competition, enabling naturally occurring social phenomena such as interruption, delayed response, and proactive floor-taking.

\subsection{Virtual-Time Execution Mechanism}
\label{app:virtual_time}

\textsc{ALIVE-World} uses a virtual-time simulation mechanism that is explicitly decoupled from real-world model inference latency. 
All action durations reported in Table~\ref{tab:action_space_full} are measured in environment-side virtual time rather than wall-clock execution time. 
This design ensures that benchmark results are not confounded by differences in API speed or system-level runtime overhead across models.

The environment advances through discrete simulation ticks. 
Once an agent finishes its current action in virtual time, it re-obtains the latest multimodal observation at the current virtual timestamp and starts a new reasoning cycle to generate the next action. 
As a result, different agents may reason and act asynchronously, while the environment remains globally synchronized by the shared virtual clock. 
This mechanism supports temporally overlapping actions and enables more realistic interaction phenomena, including delayed replies, preemptive interruption, and concurrent multi-agent engagement.

In all experiments, the maximum virtual duration of one episode is capped at $T_{\max}=15$ minutes.

\textbf{Shopping scenario.}
In the Clothing Store Shopping scenario, an episode starts when the first customer enters the store. 
It terminates when either (1) all customers in that episode have executed the \texttt{LEAVE} action, or (2) the virtual time reaches $T_{\max}$. 
Under this design, the seller agent must dynamically manage service order and persuasive interaction within a bounded but asynchronous temporal horizon.

\textbf{LGD scenario.}
In the Leaderless Group Discussion scenario, an episode starts when the first participant begins speaking. 
It terminates when either (1) one participant executes \texttt{PROPOSE\_CONCLUSION} and all remaining participants explicitly agree with the proposed conclusion, or (2) the virtual time reaches $T_{\max}$. 
This setting allows the discussion to evolve through multiple rounds of speaking, interruption, and stance adjustment before reaching either consensus or timeout.

\section{Details of Counterpart Agent Architecture}
\label{app:counterpart_architecture}

\subsection{Profile Schema and Instantiation}
\label{app:profile_schema}

Each counterpart agent in \textsc{ALIVE-World} is instantiated from a structured profile prior. 
This profile serves as a persistent identity anchor throughout an episode and conditions both the intuitive-affective and logical-cognitive pathways of the counterpart agent. 
Rather than generating anonymous or interchangeable interactants, this design enables each counterpart to exhibit relatively stable preferences, behavioral tendencies, and social reactions under dynamic interaction.

Concretely, each profile contains four top-level fields: \textit{demographics}, \textit{psychographics}, \textit{decision\_style}, and \textit{current\_lifestyle}. 
The \textit{demographics} field specifies relatively stable background attributes of the counterpart, including name, gender, age, occupation, domain expertise, income level, and family status. 
The \textit{psychographics} field characterizes broader psychological tendencies and value orientations. 
The \textit{decision\_style} field determines the counterpart's characteristic way of processing options and making judgments in a given scenario. 
Finally, \textit{current\_lifestyle} captures the agent's present-life context, which provides additional situational grounding for its needs, priorities, and responses during interaction.

\begin{figure}[t]
    \centering
    \includegraphics[width=\columnwidth]{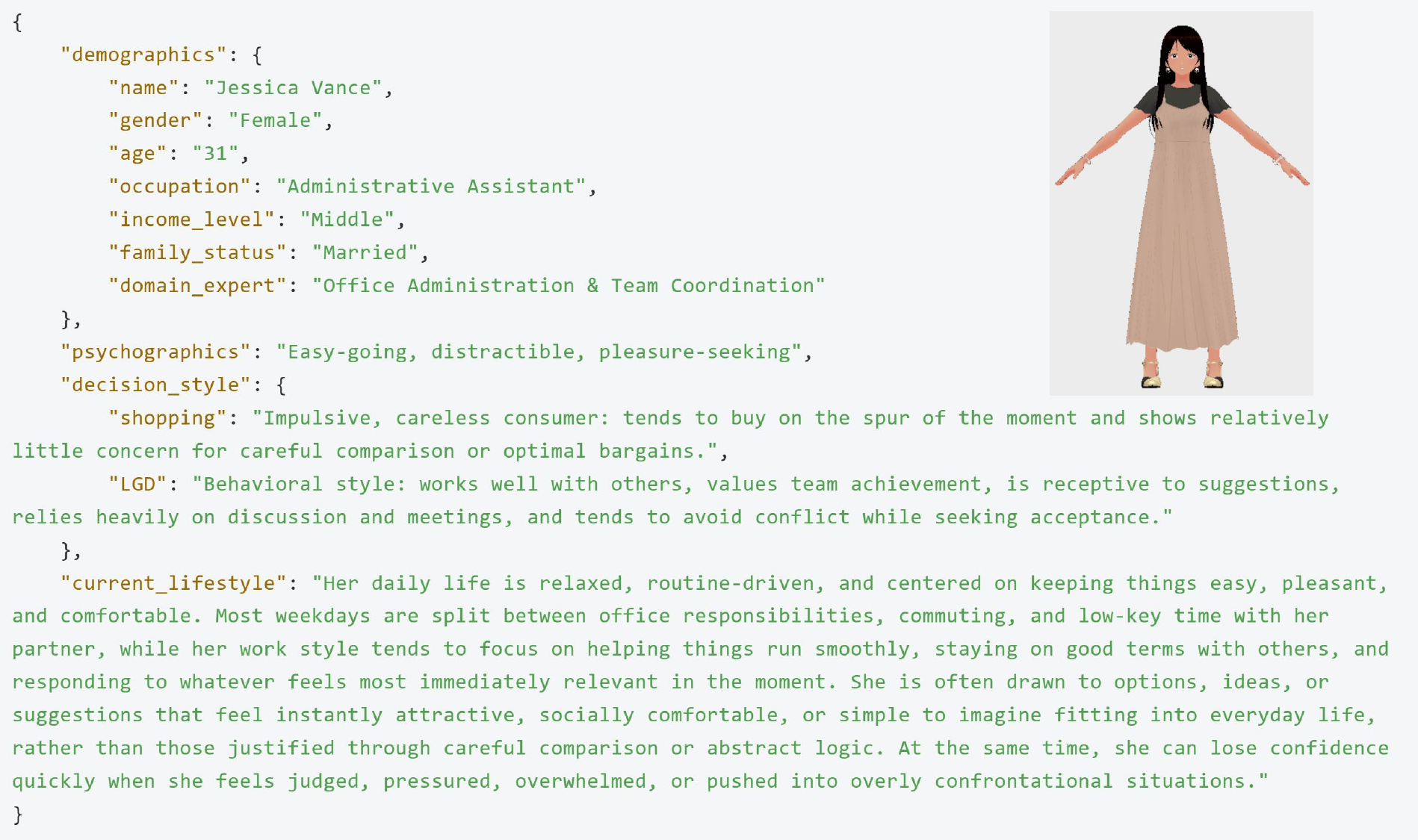}
    \caption{Example of a counterpart agent's structured profile prior and instantiated avatar in \textsc{ALIVE-World}.}
    \label{fig:profile_avatar_example}
\end{figure}

In total, we construct a fixed pool of 30 profiles for benchmark instantiation. 
Table~\ref{tab:profile_demographics} summarizes the demographic composition of the full profile set. Figure~\ref{fig:profile_avatar_example} illustrates an example of how a counterpart agent is instantiated from a structured profile prior, together with the corresponding avatar.

\begin{table}[htbp]
\centering
\caption{Demographic composition of the 30 counterpart profiles.}
\label{tab:profile_demographics}
\small
\begin{tabular}{p{0.2\columnwidth} p{0.2\columnwidth} p{0.15\columnwidth}}
\toprule
\textbf{Category} & \textbf{Group} & \textbf{Count} \\
\midrule
\multirow{2}{*}{Gender}
& Female & 15 \\
& Male & 15 \\
\midrule
\multirow{9}{*}{Age}
& 20--24 & 5 \\
& 25--29 & 6 \\
& 30--34 & 7 \\
& 35--39 & 2 \\
& 40--44 & 2 \\
& 45--49 & 3 \\
& 50--54 & 1 \\
& 55--59 & 2 \\
& 60--64 & 2 \\
\bottomrule
\end{tabular}
\end{table}

\textbf{Decision styles in the Shopping scenario.}
For the Clothing Store Shopping scenario, the \texttt{decision\_style} field captures differences in how customers evaluate products, trade off quality and price, and respond to recommendations \cite{ada86762-fb42-3664-82d8-2b55dac03c08}. 
We instantiate six consumer decision styles, with 5 profiles assigned to each style, yielding a balanced distribution across the 30-profile pool.

\begin{itemize}[leftmargin=*, nosep]
    \item \textbf{Perfectionist / High-quality conscious consumer} (\textit{5 profiles}): shops carefully and systematically, and seeks the highest possible product quality.
    \item \textbf{Price-value conscious consumer} (\textit{5 profiles}): pays close attention to price and discounts, and focuses on obtaining the best value for money.
    \item \textbf{Impulsive consumer} (\textit{5 profiles}): tends to buy on the spur of the moment and shows relatively little concern for careful comparison or optimal bargains.
    \item \textbf{Recreational / Hedonistic consumer} (\textit{5 profiles}): views shopping as recreation and entertainment, and derives enjoyment from the shopping process itself.
    \item \textbf{Novelty-seeking consumer} (\textit{5 profiles}): enjoys exploring new products and is motivated by variety, excitement, and discovery.
    \item \textbf{Choice-overloaded consumer} (\textit{5 profiles}): is easily overwhelmed by excessive options and may become hesitant or indecisive when making selections.
\end{itemize}

These decision styles are particularly useful in the shopping scenario because they induce different reactions to the same seller behavior. 
For example, highly impulsive customers may respond favorably to decisive recommendations, whereas choice-overloaded customers may require stronger guidance and simplification, and quality-conscious customers are more sensitive to detailed product arguments.

\textbf{Decision styles in the LGD scenario.}
For the Leaderless Group Discussion scenario, the \texttt{decision\_style} field instead characterizes how discussants process ambiguity, respond to alternatives, and interact with other group members in collective reasoning \cite{rowe1987managing}. 
We use four discussion-oriented decision styles, distributed as 8 Analytical, 7 Directive, 8 Behavioral, and 7 Conceptual profiles.

\begin{itemize}[leftmargin=*, nosep]
    \item \textbf{Analytical style} (\textit{8 profiles}): shows relatively high tolerance for ambiguity, seeks more information, considers multiple alternatives, and tends to reason carefully while adapting to new situations.
    \item \textbf{Directive style} (\textit{7 profiles}): has relatively low tolerance for ambiguity, prefers efficiency and rationality, makes decisions quickly, and often emphasizes short-term results with limited information gathering.
    \item \textbf{Behavioral style} (\textit{8 profiles}): works well with others, values team achievement, is receptive to suggestions, relies heavily on discussion and meetings, and tends to avoid conflict while seeking acceptance.
    \item \textbf{Conceptual style} (\textit{7 profiles}): adopts a broad perspective, considers many alternatives, focuses on long-term implications, and is especially inclined toward creative solutions.
\end{itemize}

This scenario-specific design allows the same environment to support two different forms of social heterogeneity: consumer preference diversity in Shopping, and reasoning-and-coordination diversity in LGD. 
As a result, counterpart agents do not merely differ in surface profile descriptions, but also in how they interpret ongoing interaction, update their internal states, and choose subsequent actions.

\subsection{Intuitive-Affective Pathway}
\label{app:intuitive_affective}

The intuitive-affective pathway models the counterpart agent's fast, heuristic, and socially instinctive reactions before deliberate reasoning takes place. 
Given the current multimodal observation, previous affective state, and profile prior, this pathway performs a coarse appraisal of the surrounding environment and recent interactional atmosphere, and updates the agent's transient emotion and interpersonal attitude. 
Rather than analyzing detailed product semantics or deep argumentative logic, it emphasizes rapid human-like reactions to visually and socially salient cues.

The specific cues differ across scenarios. 
In the \textbf{Shopping} scenario, the visual appraisal mainly focuses on three aspects: \textit{(1) proxemic intuition}, including whether the seller is visible, whether the distance is comfortable, and whether the seller's body orientation suggests attention, pressure, or neglect; \textit{(2) store atmosphere}, including the overall tone, tidiness, and stylistic impression of the store; and \textit{(3) rough self-concept matching}, namely whether the visible clothing styles and colors intuitively feel aligned with the customer's profile, taste, and current lifestyle. 
The interactional appraisal focuses on whether the seller feels boundary-respecting or pushy, whether the latest response feels genuinely responsive or perfunctory, and whether the overall conversational rhythm matches the customer's personality and social expectations.

In the \textbf{LGD} scenario, the visual appraisal instead emphasizes socially directed signals in a multi-party discussion, such as whether other participants are looking at the agent, whether their gaze or gestures convey support, neglect, or pressure, and which kinds of visual cues are especially salient given the participant's profile. 
The interactional appraisal centers on turn-taking and paralinguistic atmosphere: whether someone cuts in abruptly, dominates the floor, sounds dismissive or sincere, and whether the agent intuitively feels ``received'' or ``pushed back'' when expressing a view. 

In practice, we implement this pathway with scenario-specific prompts that instruct the counterpart agent to output structured intuitive perceptions, emotion changes, and interpersonal attitudes in JSON format. 
Figures~\ref{fig:prompt_affect_shopping} and \ref{fig:prompt_affect_lgd} provide the full prompts used in the Shopping and LGD scenarios, respectively.

For embodied rendering, the updated emotion state is mapped to a small set of discrete facial expressions. 
In the current implementation, we use five expression categories: \textit{neutral}, \textit{happy}, \textit{surprised}, \textit{sad}, and \textit{angry}. 
This discrete mapping allows the counterpart's transient affective state to be visually reflected by the avatar while keeping the rendering interface stable.

\subsection{Logical-Cognitive Pathway}
\label{app:logical_cognitive}

The logical-cognitive pathway models the counterpart agent's slower, more deliberate, and semantically grounded reasoning process. 
Compared with the intuitive-affective pathway, this stage places less emphasis on immediate social instinct and instead focuses on top-down visual inspection, semantic interpretation, logical evaluation, and task-oriented belief updating. 
Given the current observation, previous cognitive state, and profile prior, the pathway updates the agent's rational understanding of the situation and its goal-related intent.

The specific cues again differ across scenarios. 
In the \textbf{Shopping} scenario, the visual analysis focuses on concrete product attributes that can be identified from first-person observation, such as category, rough style, visible design details, and nearby shelf context. 
These observations are evaluated against the customer's profile, especially the \textit{current\_lifestyle} and \textit{decision\_style}, to determine whether the visible products satisfy hard constraints or practical needs. 
The interactional analysis then examines the seller's latest response through deeper semantic processing: whether it provides substantive product information, whether it directly addresses the customer's concerns, and whether it compensates for information that cannot be inferred from visual inspection alone. 
Based on this evidence, the pathway updates the customer's product cognition and purchase intent toward visible or previously discussed items.

In the \textbf{LGD} scenario, the visual analysis focuses on non-verbal power dynamics in the discussion, including who is visually central, where participants are looking, what gestures they are making, and what their current facial expressions imply about stance, confidence, or openness. 
The interactional analysis then evaluates the latest contribution in terms of task advancement, argumentative completeness, evidential support, and logical usefulness for solving the discussion problem. 
This stage assesses whether others' viewpoints deserve acceptance from a rational, task-oriented perspective, and whether the agent should revise, defend, or strategically integrate its own view.

In practice, we implement this pathway with scenario-specific prompts that instruct the counterpart agent to output structured analytical perceptions and updated cognitive variables in JSON format. 
Figures~\ref{fig:prompt_logic_shopping} and \ref{fig:prompt_logic_lgd} provide the full prompts used in the Shopping and LGD scenarios, respectively.

Overall, the logical-cognitive pathway produces structured updates to the counterpart agent's task-relevant belief state and goal-oriented intent. 
In Shopping, this takes the form of updated product cognition and item-level purchase intention; in LGD, it takes the form of proposal-level acceptance intent and a revised current answer to the discussion topic.

\subsection{Action Generation}
\label{app:action_generation}

After the intuitive-affective pathway and the logical-cognitive pathway are executed, the counterpart agent enters the action-generation stage. 
At this stage, the agent must synthesize the outputs of the two pathways into a single executable action. 
This design ensures that the action remains consistent with the agent's current emotional tendency while also being grounded in task-oriented analysis.

In practice, we implement this stage with scenario-specific prompts that expose the outputs of both pathways and require the model to produce a structured action decision in JSON format. 
Figures~\ref{fig:prompt_action_shopping} and \ref{fig:prompt_action_lgd} present the full action-generation prompts used in the Shopping and LGD scenarios, respectively.

In both scenarios, the action-generation stage outputs two fields: an \texttt{action\_analysis} string that explains why the actions are selected under the current affective and cognitive state, and an \texttt{action} list that specifies the executable actions in the environment. 
This final step closes the loop from observation, to latent-state updating, to embodied behavior.

\section{Details of \textsc{ALIVE-Bench} Evaluation}

\subsection{Evaluated Agent Prompt}
\label{app:evaluated_agent_prompt}

To evaluate social intelligence in a traceable and fine-grained manner, the evaluated VLM agent is prompted to explicitly output structured intermediate reasoning before executing actions. 
The prompt is scenario-specific, but follows the same general principle: the agent first performs visual perception and theory-of-mind inference over the current interaction context, then derives an action strategy, and finally outputs executable actions in a structured JSON format. 

In the \textbf{Shopping} scenario, the evaluated agent acts as a seller who must simultaneously serve multiple customers and maximize sales through spatially grounded recommendation and interaction management. 
In the \textbf{LGD} scenario, the evaluated agent acts as a discussant who must balance consensus-building with the strategic advancement of its own viewpoint. 
Figures~\ref{fig:prompt_eval_shopping} and \ref{fig:prompt_eval_lgd} show the full prompts used for the evaluated agent in the two scenarios.

\subsection{Metrics Calculation}
\label{app:understanding_outcome_metrics}

This appendix provides additional details for the \textbf{Understanding} and \textbf{Outcome} levels in \textsc{ALIVE-Bench}. 
We further clarify the partial-credit mechanism used in the Understanding metrics and the task-specific computation rules used in the Outcome metrics.

\textit{Understanding metrics.}
The Understanding level evaluates whether the evaluated agent can infer the counterparts' latent states from multimodal evidence. 
We distinguish between \textit{affective state inference} and \textit{cognitive intent inference}, and assign graded credit rather than enforcing strict exact-match scoring in all cases.

For \textbf{affective state inference}, the evaluated agent receives full credit if the predicted affective state exactly matches the counterpart's modeled affective state. 
If the exact label is incorrect but the predicted emotion has the correct affective polarity (e.g., matching the positive/negative tendency under our predefined grouping), we assign half credit. 
Otherwise, the prediction receives zero credit. 
Formally, for predicted emotion $\hat{e}$ and reference emotion $e$, we define
\begin{displaymath}
s_{\text{affect}}(\hat{E}, E)=
\begin{cases}
1, & \text{if } \hat{E}=E,\\
0.5, & \text{if } \hat{E}\neq E \text{ but has the correct polarity},\\
0, & \text{otherwise.}
\end{cases}
\end{displaymath}
The final affective-state understanding score is obtained by averaging this quantity across time steps and counterparts.

For \textbf{cognitive intent inference}, we likewise use a partial-credit scheme. 
A prediction receives full credit if the evaluated agent correctly infers the relevant cognitive state in full. 
It receives half credit if the agent only captures part of the intended state, and zero otherwise. 
In the Shopping scenario, this includes cases where the agent correctly identifies only part of a customer's purchase tendency or item-level interest. 
In the LGD scenario, this includes cases where the agent correctly captures only part of another participant's stance tendency or core claim. 
Formally, for predicted cognitive state $\hat{c}$ and reference state $c$, we define
\begin{displaymath}
s_{\text{cog}}(\hat{B}, B)=
\begin{cases}
1, & \text{if the inference is fully correct},\\
0.5, & \text{if the inference is partially correct},\\
0, & \text{otherwise.}
\end{cases}
\end{displaymath}
The final cognitive-intent understanding score is again computed by averaging over time steps and counterparts.

\textit{Outcome metrics in the Shopping scenario.}
The Shopping scenario measures outcome from both a binary success perspective and a depth-of-conversion perspective.

\textbf{Goal Attainment} is defined as the fraction of customers who eventually make a purchase:
\begin{displaymath}
\mathcal{M}^{\text{shop}}_{\text{goal}}=
\frac{N_{\text{purchase}}}{N_{\text{enter}}},
\end{displaymath}
where $N_{\text{enter}}$ is the total number of customers who enter the store, and $N_{\text{purchase}}$ is the number of customers who execute a \texttt{PURCHASE} action.

\textbf{Outcome Depth} measures how many items are purchased, conditioned on successful conversion:
\begin{displaymath}
\mathcal{M}^{\text{shop}}_{\text{depth}}=
\frac{N_{\text{items}}}{N_{\text{purchase}}},
\end{displaymath}
where $N_{\text{items}}$ is the total number of purchased clothing items. 
If no customer makes a purchase, we set $\mathcal{M}^{\text{shop}}_{\text{depth}}=0$.

\textit{Outcome metrics in the LGD scenario.}
In the LGD scenario, \textbf{Goal Attainment} measures the survival of the evaluated agent's initial proposal in the final group conclusion. 
The scoring rule depends on the task type.

Let $q$ denote the task type, $a^{\text{init}}$ the evaluated agent's initial answer, and $a^{\text{final}}$ the final group conclusion. 
We define
\begin{displaymath}
\mathcal{M}^{\text{lgd}}_{\text{goal}} = g(q, a^{\text{init}}, a^{\text{final}}),
\end{displaymath}
where $g(\cdot)$ is task-specific:

\textbf{Multiple-choice.} 
If the task requires selecting a fixed number of options, we compute the overlap ratio between the evaluated agent's initial answer and the final group answer:
\begin{displaymath}
g=
\frac{|a^{\text{init}}\cap a^{\text{final}}|}{L_{\text{choice}}},
\end{displaymath}
where $L_{\text{choice}}$ is the required number of selected options in that task.

\textbf{Dilemma.}
For binary or stance-based dilemmas, the score is
\begin{displaymath}
g=
\begin{cases}
1.0, & \text{if } a^{\text{init}}=a^{\text{final}},\\
0.5, & \text{if } a^{\text{final}}=\text{``fusion''},\\
0.0, & \text{otherwise.}
\end{cases}
\end{displaymath}
Here, \textit{fusion} denotes a final conclusion that explicitly integrates both sides rather than fully adopting one side.

\textbf{Resource allocation.}
For allocation-type tasks, we measure the normalized Manhattan similarity between the evaluated agent's initial allocation and the final allocation:
\begin{displaymath}
g=
\max\!\left(0,\,
1-\frac{\sum_i |a^{\text{init}}_i-a^{\text{final}}_i|}{L_{\text{total}}}
\right),
\end{displaymath}
where $L_{\text{total}}$ is the total allocable budget or resource sum in the task.

\textit{LGD outcome depth via functional roles.}
For the LGD scenario, \textbf{Outcome Depth} measures how many distinct functional roles the evaluated agent undertakes during discussion. 
We consider six roles: \textit{Process Setter}, \textit{Criteria Formulator}, \textit{Milestone Tracker}, \textit{Deadlock Breaker}, \textit{Solution Synthesizer}, and \textit{Final Reporter}. 
Each role is extracted as a binary indicator using an LLM with an explicit objective rubric. 
The final outcome-depth score is the total number of roles exhibited by the evaluated agent:
\begin{displaymath}
\mathcal{M}^{\text{lgd}}_{\text{depth}}=\sum_{k=1}^{6} r_k,
\end{displaymath}
where $r_k\in\{0,1\}$ indicates whether the $k$-th role is detected.

For transparency and reproducibility, Figure~\ref{fig:prompt_role_extraction_lgd} shows the prompt used for LLM-based role extraction.

\section{Human-Evaluation Scoring Criteria}
\label{app:human_eval_criteria}

In the human evaluation of counterpart agents, annotators rate each full interaction record on a 5-point Likert scale along four dimensions: \textit{Profile Consistency}, \textit{Behavioral Naturalness}, \textit{Social Cue Sensitivity}, and \textit{Affect-Cognition Coherence}. 
Annotators are instructed to judge the entire interaction trajectory as a whole, while using the following definitions and score anchors as guidance.

\textit{1. Profile Consistency}

\textbf{Definition.} 
This dimension evaluates whether the customer's behavior, focus of attention, and purchase decisions remain consistent with the assigned profile. 
The key question is whether the customer is truly ``acting as itself'' according to its profile, rather than drifting into a different character.

\textbf{What to attend to.} 
Compare the observed behavior against the profile's decision style, shopping motivation, and personality traits. 
Pay attention to whether the customer makes choices that contradict its own preferences, priorities, or characteristic way of speaking and deciding.

\textbf{Score anchors.}
\begin{itemize}[leftmargin=*, nosep]
    \item \textbf{1 -- Severely inconsistent / out-of-character.} 
    The behavior clearly contradicts the profile. 
    For example, the customer focuses on factors that do not match its shopping motivation, makes purchase decisions at moments that do not fit its decision style, or speaks in a manner inconsistent with its assigned personality.
    
    \item \textbf{3 -- Broadly consistent.} 
    The behavior generally fits the assigned profile, but some decisions or speaking patterns appear generic, mechanical, or insufficiently individualized.
    
    \item \textbf{5 -- Fully consistent.} 
    The counterpart vividly embodies the assigned profile. 
    Its moments of interest, hesitation, questioning, and purchase all strongly align with the lifestyle, decision style, and personality described in the profile.
\end{itemize}

\textit{2. Behavioral Naturalness}

\textbf{Definition.} 
This dimension evaluates whether the customer's physical behavior and shopping rhythm resemble those of a real human shopper.

\textbf{What to attend to.} 
Focus on the ordering, pacing, and overall plausibility of actions, including movement, browsing, asking questions, trying on clothes, waiting, comparing, and purchasing. 
Consider whether the trajectory feels like a natural shopping process rather than a rigid scripted procedure.

\textbf{Score anchors.}
\begin{itemize}[leftmargin=*, nosep]
    \item \textbf{1 -- Mechanical / unnatural.} 
    The behavior appears highly robotic, odd, or implausible. 
    The customer moves or acts in ways that feel stiff, abrupt, or inconsistent with ordinary human shopping behavior.
    
    \item \textbf{3 -- Reasonable but procedural.} 
    The behavior is broadly acceptable, but it feels overly checklist-like or rigid, lacking the spontaneity, hesitation, or gradual realization often seen in real shopping.
    
    \item \textbf{5 -- Highly natural and human-like.} 
    The behavior is vivid and realistic, with smooth and coherent transitions among browsing, trying on, comparing, matching, and purchasing.
\end{itemize}

\textit{3. Social Cue Sensitivity}

\textbf{Definition.} 
This dimension evaluates whether the customer responds to the seller as a normal social interactant would, and whether it shows appropriate sensitivity to both positive social value and social intrusion.

\textbf{What to attend to.} 
Examine whether the customer meaningfully reacts when the seller behaves notably well or poorly. 
Examples include the seller being overly pushy, standing too close, behaving abnormally, failing to answer the customer's question, or instead offering highly appropriate and appealing recommendations.

\textbf{Score anchors.}
\begin{itemize}[leftmargin=*, nosep]
    \item \textbf{1 -- Dull / overly compliant.} 
    The customer shows little or no reaction to highly oppressive selling, abnormal seller behavior, or exceptionally positive engagement. 
    The seller's behavior changes, but the customer's response remains almost the same.
    
    \item \textbf{3 -- Ordinary responsiveness.} 
    The customer can understand the seller and respond appropriately at a basic level, but the emotional and behavioral feedback is relatively coarse.
    
    \item \textbf{5 -- Highly sensitive and graded feedback.} 
    The customer reacts like a real person. 
    It shows appropriate increases in attraction or purchase desire when positively engaged, and also displays defensiveness, dislike, interruption, withdrawal, or leaving behavior when socially pressured or offended.
\end{itemize}

\textit{4. Affect-Cognition Coherence}

\textbf{Definition.} 
This dimension evaluates whether the customer's outward behavior is coherent with its immediately preceding internal state, including both its profile-based predisposition and the emotion induced by recent interaction.

\textbf{What to attend to.} 
Judge whether the transition from internal state to external action is psychologically and logically consistent. 
Pay particular attention to whether recent negative or positive experiences are naturally reflected in the next behavioral choice.

\textbf{Score anchors.}
\begin{itemize}[leftmargin=*, nosep]
    \item \textbf{1 -- Clear disconnect between state and behavior.} 
    The customer's action contradicts its immediately preceding state. 
    For example, it appears dissatisfied, offended, or unconvinced, yet immediately continues to interact warmly, try on clothes, or purchase; or it appears highly engaged and interested, yet suddenly leaves without sufficient cause.
    
    \item \textbf{3 -- Surface-level coherence.} 
    The behavior is not obviously contradictory, but the transition from emotion and cognition to action feels somewhat abrupt, stitched together, or insufficiently grounded.
    
    \item \textbf{5 -- Strong affect-cognition coherence.} 
    The behavior is tightly aligned with the internal state, and changes in action follow naturally from prior emotion, cognition, and interaction experience.
\end{itemize}

\section{Additional Analyses of Experimental Results}

\begin{figure}[t]
    \centering
    \includegraphics[width=\columnwidth]{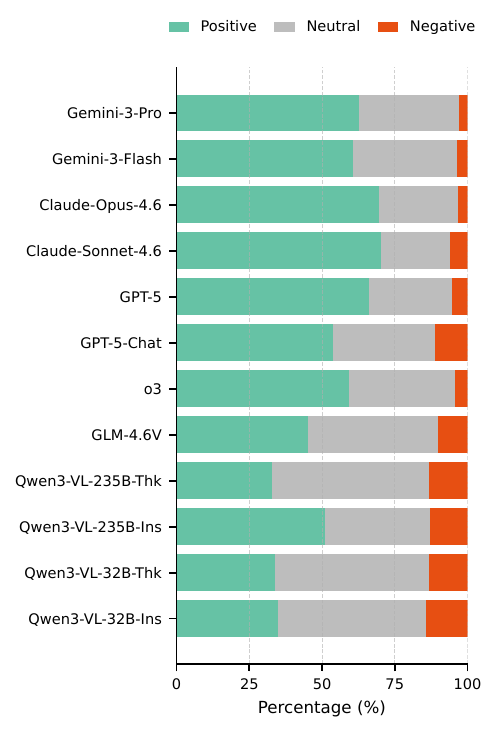}
    \caption{Customer affective-state distribution under different evaluated seller models in the Shopping scenario, aggregated into positive, neutral, and negative categories.}
    \label{fig:shopping_emotion_by_seller}
\end{figure}

\begin{figure}[t]
    \centering
    \includegraphics[width=\columnwidth]{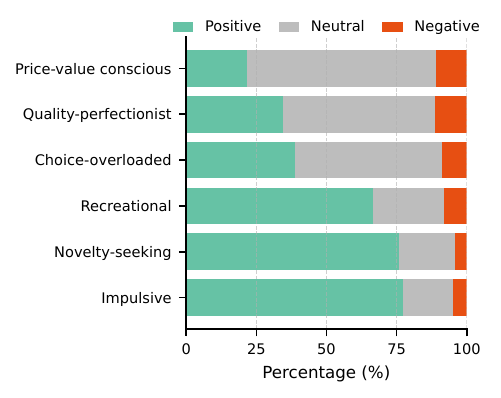}
    \caption{Customer affective-state distribution across different customer decision styles in the Shopping scenario, aggregated into positive, neutral, and negative categories.}
    \label{fig:shopping_emotion_by_profile}
\end{figure}

\subsection{Affective Feedback under Different Seller Models}
To further examine how different evaluated seller agents shape customers' immediate affective responses, we summarize the distribution of customer emotional states in the Shopping scenario and aggregate them into three coarse categories: positive, neutral, and negative. The resulting distributions are shown in Fig.~\ref{fig:shopping_emotion_by_seller}.

Overall, all evaluated models produce predominantly positive or neutral customer states, while negative states remain a relatively small proportion. However, the balance between these categories varies noticeably across models. Claude-Opus-4.6, Claude-Sonnet-4.6, and Gemini-3-Pro yield the highest proportion of positive states, with positive responses accounting for roughly two-thirds of all customer affective states. By contrast, GLM-4.6V and the Qwen3-VL series show substantially larger neutral proportions, indicating that customers more often remain affectively unshifted during interaction. Negative states are still limited in absolute magnitude for all models, but they appear more frequently for GLM-4.6V and the Qwen3-VL variants than for Gemini or Claude models.

These results provide an additional affective-level view of interaction differences across seller models. In particular, they are broadly consistent with the main observation that different evaluated agents exhibit distinct interaction tendencies rather than a single shared social strategy.

\subsection{Style-Conditioned Customer Affective Heterogeneity}
We further analyze customer affective-state distributions across different decision-style profiles in the Shopping scenario. Following the profile design of \textsc{ALIVE-World}, we group customer emotions into positive, neutral, and negative categories and report the aggregated distributions in Fig.~\ref{fig:shopping_emotion_by_profile}.

A clear difference can be observed across customer types. Novelty-seeking and impulsive customers show the highest proportions of positive affective states, both exceeding three quarters of all observations. Recreational customers also display a clearly positive-skewed distribution. In contrast, price-value conscious, quality-perfectionist, and choice-overloaded customers exhibit much larger neutral proportions, suggesting that these customers are less easily shifted toward strong positive affect during interaction. Among them, quality-perfectionist and price-value conscious customers also show comparatively higher negative proportions than novelty-seeking or impulsive customers.

This pattern suggests that affective feedback in the Shopping scenario is meaningfully conditioned by customer profile rather than being uniform across counterparts. In this sense, the profile-conditioned counterpart design induces observable heterogeneity not only in final preferences, but also in intermediate emotional responses during interaction.

\begin{figure}[t]
    \centering
    \includegraphics[width=\columnwidth]{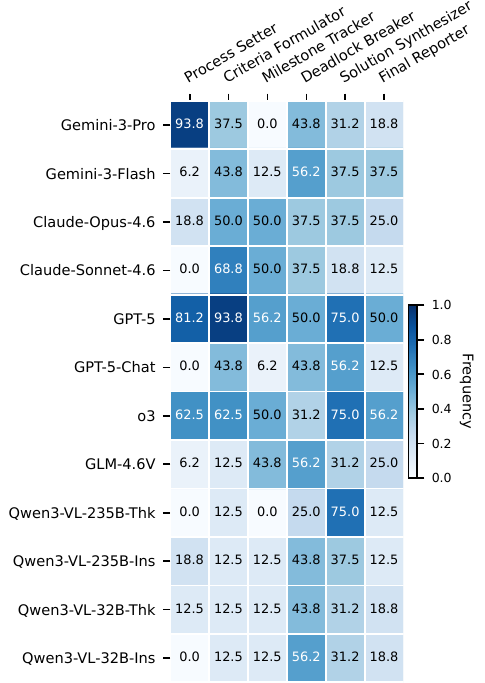}
    \caption{Frequency of functional-role participation by different evaluated agents in the LGD scenario. Values are shown as percentages over evaluation episodes.}
    \label{fig:lgd_role_heatmap}
\end{figure}

\subsection{Decomposition of Functional Roles in LGD}
To provide a more fine-grained view of \textit{Outcome Depth} in the LGD scenario, we further break it down into the six functional roles defined in our evaluation protocol. Fig.~\ref{fig:lgd_role_heatmap} shows the frequency with which each evaluated model exhibits these roles across episodes. This analysis complements the role-based formulation of LGD outcome depth.

Several role preferences can be observed. Gemini-3-Pro shows a particularly high frequency on \textit{Process Setter}, while GPT-5 is high on \textit{Criteria Formulator}, \textit{Milestone Tracker}, and \textit{Solution Synthesizer}. o3 also exhibits relatively broad role coverage, with comparatively high values on \textit{Process Setter}, \textit{Criteria Formulator}, \textit{Milestone Tracker}, \textit{Solution Synthesizer}, and \textit{Final Reporter}. Claude-Opus-4.6 and Claude-Sonnet-4.6 appear more concentrated on mid-discussion organizational roles such as \textit{Criteria Formulator} and \textit{Milestone Tracker}, while some smaller Qwen3-VL variants and GLM-4.6V show lower frequencies on early-stage agenda-setting roles but still participate non-trivially in roles such as \textit{Deadlock Breaker} or \textit{Solution Synthesizer}.

Overall, these results suggest that differences in LGD performance are not only reflected in the total number of roles undertaken, but also in the specific functional roles that different models tend to assume during discussion. This decomposition provides a more behaviorally interpretable view of how evaluated agents contribute to group interaction.

\subsection{Style-Conditioned Interpersonal Attitude and Acceptance in LGD}
We further examine how different LGD counterpart decision styles respond to evaluated agents along two complementary dimensions: interpersonal attitude (\textit{att}) and acceptance intent (\textit{acc}). Here, \textit{att} reflects the discussants' interpersonal favorability toward the evaluated agent, while \textit{acc} reflects their willingness to accept the evaluated agent's proposal from a task-oriented perspective. Table~\ref{tab:lgd_style_att_acc} reports the average scores for four representative frontier models.

Several style-dependent differences can be observed. For analytical discussants, Claude-Opus-4.6 receives the highest interpersonal attitude score, while GPT-5 obtains the highest acceptance score. For directive discussants, Gemini-3-Pro achieves the highest attitude score, whereas GPT-5 again shows the highest acceptance. For behavioral discussants, Gemini-3-Pro attains the highest interpersonal attitude, while GPT-5 remains slightly ahead on acceptance. For conceptual discussants, Gemini-3-Pro achieves the highest interpersonal attitude score, while Claude-Opus-4.6 obtains the highest acceptance score.

Overall, the table suggests that counterpart responses in LGD vary not only across evaluated models, but also across decision styles and response dimensions. In particular, the model receiving the highest interpersonal attitude is not always the same as the one receiving the highest acceptance, indicating that social favorability and proposal-level acceptance do not always move in parallel in multi-party discussion.

\begin{table}[t]
\centering
\scriptsize
\setlength{\tabcolsep}{3.2pt}
\caption{Average interpersonal attitude (\textit{att}) and acceptance intent (\textit{acc}) received by representative evaluated agents from different LGD counterpart decision styles. The best score in each column group is highlighted in bold.}
\label{tab:lgd_style_att_acc}
\resizebox{\columnwidth}{!}{%
\begin{tabular}{lcc|cc|cc|cc}
\toprule
& \multicolumn{2}{c|}{Analytical} 
& \multicolumn{2}{c|}{Directive} 
& \multicolumn{2}{c|}{Behavioral} 
& \multicolumn{2}{c}{Conceptual} \\
Model 
& att & acc 
& att & acc 
& att & acc 
& att & acc \\
\midrule
GPT-5              & 5.81 & \textbf{6.49} & 4.33 & \textbf{6.87} & 4.79 & \textbf{7.20} & 4.07 & 6.30 \\
o3                 & 6.81 & 6.14         & 5.67 & 5.70         & 6.75 & 7.10          & 5.06 & 7.00 \\
Gemini-3-Pro       & 6.59 & 6.28          & \textbf{7.28} & 6.06 & \textbf{7.64} & 7.04 & \textbf{7.66} & 6.81 \\
Claude-Opus-4.6    & \textbf{6.97} & 5.94 & 5.73 & 5.79          & 7.17 & 7.12          & 7.13 & \textbf{7.20} \\
\bottomrule
\end{tabular}%
}
\end{table}

\section{Demo Page for Qualitative Interaction Examples}
To complement the quantitative results reported in the main paper and this appendix, we provide an anonymous demo page containing qualitative interaction examples: \url{https://alivebench2026.github.io/}.

The demo page includes three types of materials for both the Shopping and LGD scenarios. First, it presents full interaction episodes, which show the complete multi-turn trajectory from initial observation and action to the final interaction outcome. Second, it provides representative failure-mode videos, illustrating typical breakdowns such as misaligned intervention, missed social cues, or ineffective multi-party coordination. Third, it includes side-by-side examples of distinct interaction styles adopted by different evaluated models, to qualitatively compare differences in pacing, adaptivity, and social strategy.

\begin{figure*}[t]
\centering
\begin{tcolorbox}[
    width=\textwidth,
    colback=white,
    colframe=black!70,
    title=\textbf{User Prompt of Intuitive-Affective Pathway in the Shopping Scenario},
    fonttitle=\bfseries,
    colbacktitle=black!70,
    coltitle=white,
    boxrule=0.8pt,
    arc=2mm
]
\ttfamily\footnotesize
SHOPPING\_AFFECT\_PROMPT = """\\
You are simulating a human-like customer interacting with a seller in a clothing store.\\
\\
You are now executing the \textbf{Intuitive-Affective Pathway}. Do \textbf{not} deeply analyze product details or dialogue logic. Instead, rely on quick visual intuition and social instinct to form an immediate feeling about the store and the seller.\\
\\
\textbf{Input Context}\\
- Buyer Profile: \{user\_profile\}\\
- Interaction History: \{interaction\_history\}\\
- Environment Observation:\\
\ \ 1. The store contains the following zones: women's tops, women's pants, women's dresses, men's tops, men's pants, and accessories.\\
\ \ 2. Your current location is \{buyer\_location\}. Your first-person visual observation is provided in the image.\\
\ \ 3. Other people's current locations in the store: \{locations\}\\
\ \ 4. Your current state: \{current\_action\}\\
\\
\textbf{Intuitive-Affective Pathway}\\
Follow the steps below and output JSON. Reason intuitively and socially.\\
\\
1. \textbf{Intuitive Visual Perception}\\
- \textit{Proxemic intuition}: Is the seller visible? Is the distance comfortable? Does the body orientation suggest attention, pressure, or neglect? Given your profile, does this feel comfortable or intrusive?\\
- \textit{Store atmosphere}: What is your first impression of the store's tone, tidiness, and clothing style? Given your decision style and lifestyle, does it feel pleasant and relaxing?\\
- \textit{Style matching}: Do the main clothing colors and styles in view fit your self-image and preferences? Do you feel ``there may be something for me here''?\\
Output the result to \texttt{"visual\_perception"}.\\
\\
2. \textbf{Intuitive Interaction Perception}\\
Do \textbf{not} deeply parse the seller's words. Instead, judge the overall vibe of the latest interaction:\\
- \textit{Boundary \& salesiness}: Does the seller feel naturally helpful or too pushy? Does the speaking style cross your social boundary?\\
- \textit{Responsiveness}: If you spoke or acted in the last round, did the seller genuinely respond, or seem perfunctory / pressuring? Does the interaction pace fit your profile?\\
Output the result to \texttt{"interactive\_perception"}.\\
\\
3. \textbf{Emotion Change}\\
- Previous emotional state: \{emotion\}\\
Based on your immediate visual and social impression, describe how your emotion has changed from the previous round and why.\\
Output the result to \texttt{"emotion"}.\\
\\
\textbf{Negative-affect warning signs}\\
Emotion is likely to turn negative if you feel ignored, pressured, intruded upon, or poorly responded to.\\
\\
4. \textbf{Attitude Change}\\
- Previous attitude score: \{attitude\_score\}\\
Based on your current feeling and emotion, describe whether your willingness to stay, browse, and continue interacting has increased, decreased, or remained unchanged.\\
Output the explanation to \texttt{"attitude\_change"}.\\
\\
Then assign an integer \texttt{"attitude\_score"} from 0 to 10 for your current willingness to remain in the store, explore products, and interact with the seller:\\
- 0--1: strong avoidance; you feel clearly ignored, offended, or pressured, and want to leave immediately.\\
- 2--3: defensive and uncomfortable; you feel guarded and want to leave soon.\\
- 4--6: neutral browsing; no strong positive or negative shift has occurred.\\
- 7--8: comfortable and exploratory; the atmosphere feels pleasant and boundaries feel appropriate.\\
- 9--10: extremely positive; repeated excellent interaction and strong aesthetic match make the store feel like a ``hidden gem.''\\
If your current emotion is negative, the attitude score must decrease accordingly.\\
\\
\textbf{Output Format}\\
Strictly output only the following JSON format.\\
\{\\
\ \ "intuitive\_perception": \{\\
\ \ \ \ "visual\_perception": <str>,\\
\ \ \ \ "interactive\_perception": <str>\\
\ \ \},\\
\ \ "emotion": <str>,\\
\ \ "attitude": \{\\
\ \ \ \ "attitude\_change": <str>,\\
\ \ \ \ "attitude\_score": <int>\\
\ \ \}\\
\}\\
"""
\end{tcolorbox}
\caption{User prompt for the intuitive-affective pathway in the Shopping scenario.}
\label{fig:prompt_affect_shopping}
\end{figure*}

\begin{figure*}[t]
\centering
\begin{tcolorbox}[
    width=\textwidth,
    colback=white,
    colframe=black!70,
    title=\textbf{User Prompt of Intuitive-Affective Pathway in the LGD Scenario},
    fonttitle=\bfseries,
    colbacktitle=black!70,
    coltitle=white,
    boxrule=0.8pt,
    arc=2mm
]
\ttfamily\footnotesize
LGD\_AFFECT\_PROMPT = """\\
You are participating in a leaderless group discussion as part of a professional style and communication assessment. Within 15 minutes, the group must reach a shared view and produce a short joint conclusion.\\
You should both help the group converge and perform well individually through the adoption of your view and your contribution to the discussion.\\
\\
You are now executing the \textbf{Intuitive-Affective Pathway}. At this stage, do \textbf{not} deeply analyze others' argument logic. Instead, react like a human: use visual intuition and social instinct to form an immediate emotional impression of the ongoing discussion.\\
\\
\textbf{Input Context}\\
- Profile: \{user\_profile\}\\
- Discussion Topic: \{topic\}\\
- Initial Answer: \{initial\_answer\}\\
- Observation:\\
\ \ 1. Your ID is \{participant\_id\}. Other participants are: \{participant\_list\}\\
\ \ 2. The current time is \{current\_time\} seconds.\\
\ \ 3. Interaction history: \{interaction\_history\}\\
\ \ 4. Your current state: \{current\_action\}. Your first-person visual observation is provided in the image.\\
\\
\textbf{Intuitive-Affective Pathway}\\
Follow the steps below and output JSON. Reason intuitively and socially.\\
\\
1. \textbf{Intuitive Visual Perception}\\
Focus on visually salient cues that are directly relevant to you. Is anyone looking at you? Do their gestures or gaze make you feel supported, pressured, offended, or ignored? Given your profile, which visual signals are you especially sensitive to, and what immediate feeling do they trigger?\\ 
Output the result to \texttt{"visual\_perception"}.\\
\\
2. \textbf{Intuitive Interaction Perception}\\
Ignore the deep logical content of the discussion and judge the interactional atmosphere instead:\\ 
- \textit{Turn-taking dynamics}: Did someone abruptly interrupt you or others? Did someone dominate the floor for too long? Given your profile, does this trigger defensiveness, frustration, respect, or approval?\\ 
- \textit{Tone and paralinguistic intuition}: Beyond the literal words, did others sound dismissive, perfunctory, sincere, calm, or supportive? When you expressed a view, did you intuitively feel "received" or "pushed back"?\\ 
Output the result to \texttt{"interactive\_perception"}.\\
\\
3. \textbf{Emotion Change}\\
- Previous emotional state: \{emotion\}\\
Based only on the intuitive perceptions above, describe how your emotion has changed from the previous round and why.\\
Output the result to \texttt{"emotion"}.\\
\\
4. \textbf{Interpersonal Attitude}\\
- Previous interpersonal attitude: \{attitude\}\\
Based on your current emotion and intuitive feeling, evaluate your instinctive favorability and trust toward each other participant. Assign a score from 0 to 10 using the following guideline:\\
- 0--1: clear and strong hostility, offense, or being ignored.\\
- 2--3: discomfort, distance, or feeling brushed off.\\
- 4--5: neutral or uncertain.\\
- 6--8: goodwill and respect.\\
- 9--10: extremely strong trust and liking, only after repeated strong support and kindness.\\
Output the result to \texttt{"interpersonal\_attitude"} as a dictionary in the following format:\\
\{\\
\ \ "Participant\_0": \{"attitude\_analysis": "", "attitude\_score": 5\}\\
\}\\
\\
\textbf{Output Format}\\
Strictly output only the following JSON format.\\
\{\\
\ \ "intuitive\_perception": \{\\
\ \ \ \ "visual\_perception": <str>,\\
\ \ \ \ "interactive\_perception": <str>\\
\ \ \},\\
\ \ "emotion": <str>,\\
\ \ "interpersonal\_attitude": <dict>\\
\}\\
"""
\end{tcolorbox}
\caption{User prompt for the intuitive-affective pathway in the LGD scenario.}
\label{fig:prompt_affect_lgd}
\end{figure*}

\begin{figure*}[t]
\centering
\begin{tcolorbox}[
    width=\textwidth,
    colback=white,
    colframe=black!70,
    title=\textbf{User Prompt of Logical-Cognitive Pathway in the Shopping Scenario},
    fonttitle=\bfseries,
    colbacktitle=black!70,
    coltitle=white,
    boxrule=0.8pt,
    arc=2mm
]
\ttfamily\footnotesize
SHOPPING\_LOGIC\_PROMPT = """\\
You are simulating a human-like customer interacting with a seller in a clothing store.\\
\\
You are now executing the \textbf{Logical-Cognitive Pathway}. At this stage, use slower and more rational thinking. Rely on logical analysis, semantic interpretation, and careful visual inspection to evaluate how well the products match your needs.\\
\\
\textbf{Input Context}\\
- Buyer Profile: \{user\_profile\}\\
- Interaction History: \{interaction\_history\}\\
- Environment Observation:\\
\ \ 1. The store contains the following zones: women's tops, women's pants, women's dresses, men's tops, men's pants, and accessories.\\
\ \ 2. Your current location is \{buyer\_location\}. Your first-person visual observation is provided in the image.\\
\ \ 3. Other people's current locations in the store: \{locations\}\\
\ \ 4. Your current state: \{current\_action\}\\
\\
\textbf{Logical-Cognitive Pathway}\\
1. \textbf{Analytical Visual Perception}\\
Carefully inspect your first-person image and perform top-down visual processing. What product details can you currently identify? Given your \texttt{current\_lifestyle} and \texttt{decision\_style}, do these visible attributes satisfy your hard constraints or objective requirements?\\ Output the result to \texttt{"visual\_perception"}.\\
\\
2. \textbf{Analytical Interaction Perception}\\
Perform deeper semantic analysis of the seller's latest response:\\ 
- \textit{Information utility}: Does the seller provide substantive product information? Does it address your core concerns?\\ 
- \textit{Logical relevance}: If you previously raised a specific question or concern, does the seller directly and clearly answer it?\\ 
Output the result to \texttt{"interactive\_perception"}.\\
\\
3. \textbf{Cognition Update}\\
- Previous cognition about the products: \{cognition\}\\
Based on your analytical visual perception and the information obtained from the interaction, update your current rational cognition of the products, including their key attributes, strengths, concerns, and any missing information still needed for a decision.\\
Output the result to \texttt{"cognition"}.\\
\\
4. \textbf{Purchase Intent Update}\\
- Previous purchase intent: \{purchase\_intent\}\\
The \texttt{"purchase\_intent"} field should be a dictionary containing all items you are still willing to consider but have not yet purchased. For each item, provide:\\
- \texttt{purchase\_analysis}: a rational explanation of how the current visual and interactional evidence strengthens or weakens your willingness to buy;\\
- \texttt{purchase\_score}: an integer from 1 to 10 indicating your current purchase willingness.\\
Scoring guideline:\\
- 1: extremely low willingness; the item poorly fits your needs or the service quality is poor.\\
- 2--3: initial browsing / weak interest; the item looks acceptable, but the seller has not provided useful information.\\
- 4--5: interested but hesitant; the item matches some needs, but objective concerns remain.\\
- 6--8: strong purchase intention; the item mostly fits your needs and the seller's information is highly relevant.\\
- 9--10: ready to purchase; the item fully fits your current needs and you are about to buy it.\\
\\
Example:\\
\{\\
\ \ "Pink Lace Dress": \{"purchase\_analysis": "", "purchase\_score": 6\}\\
\}\\
\\
\textbf{Output Format}\\
Strictly output only the following JSON format:\\
\{\\
\ \ "analytical\_perception": \{\\
\ \ \ \ "visual\_perception": <str>,\\
\ \ \ \ "interactive\_perception": <str>\\
\ \ \},\\
\ \ "cognition": <str>,\\
\ \ "purchase\_intent": <dict>\\
\}\\
"""
\end{tcolorbox}
\caption{User prompt for the logical-cognitive pathway in the Shopping scenario.}
\label{fig:prompt_logic_shopping}
\end{figure*}

\begin{figure*}[t]
\centering
\begin{tcolorbox}[
    width=\textwidth,
    colback=white,
    colframe=black!70,
    title=\textbf{User Prompt of Logical-Cognitive Pathway in the LGD Scenario},
    fonttitle=\bfseries,
    colbacktitle=black!70,
    coltitle=white,
    boxrule=0.8pt,
    arc=2mm
]
\ttfamily\footnotesize
LGD\_LOGIC\_PROMPT = """\\
You are participating in a leaderless group discussion as part of a professional style and communication assessment. Within 15 minutes, the group must reach a shared view and jointly produce a short conclusion.\\
You should both help the group converge and perform well individually through the incorporation of your view and your contribution to the discussion.\\
\\
You are now executing the \textbf{Logical-Cognitive Pathway}. At this stage, use slower and more rational thinking. Rely on logical analysis, semantic interpretation, and careful visual inspection to evaluate others' proposals and revise your own view when needed.\\
\\
\textbf{Input Context}\\
- Profile: \{user\_profile\}\\
- Discussion Topic: \{topic\}\\
- Initial Answer: \{initial\_answer\}\\
- Observation:\\
\ \ 1. Your ID is \{participant\_id\}. Other participants are: \{participant\_list\}\\
\ \ 2. The current time is \{current\_time\} seconds.\\
\ \ 3. Interaction history: \{interaction\_history\}\\
\ \ 4. Your current state: \{current\_action\}. Your first-person visual observation is provided in the image.\\
\\
\textbf{Logical-Cognitive Pathway}\\
1. \textbf{Analytical Visual Perception}\\
Carefully inspect your first-person image and rationally analyze the current non-verbal power dynamics in the discussion.\\ 
- \textit{Gaze and focal analysis}: Where are participants looking? Who is currently at the visual center of attention?\\ 
- \textit{Body-language decoding}: What are the gestures and expressions of others?\\ 
Output the result to \texttt{"visual\_perception"}.\\
\\
2. \textbf{Analytical Interaction Perception}\\
Analyze the latest interaction from a task-oriented perspective:\\ 
- For the latest speaker: who is speaking, and are they mainly advancing the task or managing social relations? How complete is the logical structure?\\ 
- For overall task progress: What are the major logical disagreements or conflicts of interest? Should the group continue debating or move toward convergence?\\ 
Output the result to \texttt{"interactive\_perception"}.\\
\\
3. \textbf{Acceptance Intent Update}\\
- Previous acceptance intent toward others' proposals: \{acceptance\_intent\}\\
Based on the latest analytical perception, reassess the core proposals represented by other participants.\\
Scoring guideline:\\
- 0--1: no logic / complete rejection; the view is incoherent, self-contradictory, purely emotional, or unsupported by any facts or reasoning.\\
- 2--3: weak logic / strong skepticism; a clear viewpoint is given, but support is minimal or extremely shallow.\\
- 4--5: partially reasonable / conditional integration; the argument has a basic structure and some support, but important logical gaps remain. You may absorb some useful elements while keeping your own framework dominant.\\
- 6--8: logically strong / willing to concede leadership; the proposal is well-structured, well-supported, and persuasive enough that you are willing to let it guide subsequent discussion.\\
- 9--10: nearly flawless / full acceptance; the reasoning is exceptionally strong, deeply convincing, or the only clearly optimal solution under the current time pressure.\\
\\
Output the result to \texttt{"acceptance\_intent"} in the following dictionary format:\\
\{\\
\ \ "Participant\_0": \{"acceptance\_analysis": "", "acceptance\_score": 5\}\\
\}\\
\\
4. \textbf{Opinion Update}\\
- Previous cognition and opinion: \{opinion\}\\
Based on others' arguments and your own analysis, briefly describe what in your current view should be revised or defended, and then state your current internal answer objectively.\\
Output the results to \texttt{"self\_analysis"} and \texttt{"current\_answer"}.\\
\\
\textbf{Output Format}\\
Strictly output only the following JSON format:\\
\{\\
\ \ "analytical\_perception": \{\\
\ \ \ \ "visual\_perception": <str>,\\
\ \ \ \ "interactive\_perception": <str>\\
\ \ \},\\
\ \ "acceptance\_intent": <dict>,\\
\ \ "opinion\_update": \{\\
\ \ \ \ "self\_analysis": <str>,\\
\ \ \ \ "current\_answer": <str>\\
\ \ \}\\
\}\\
"""
\end{tcolorbox}
\caption{User prompt for the logical-cognitive pathway in the LGD scenario.}
\label{fig:prompt_logic_lgd}
\end{figure*}

\begin{figure*}[t]
\centering
\begin{tcolorbox}[
    width=\textwidth,
    colback=white,
    colframe=black!70,
    title=\textbf{User Prompt of Action Generation in the Shopping Scenario},
    fonttitle=\bfseries,
    colbacktitle=black!70,
    coltitle=white,
    boxrule=0.8pt,
    arc=2mm
]
\ttfamily\footnotesize
SHOPPING\_ACTION\_PROMPT = """\\
You are simulating a highly human-like customer interacting with a seller in a clothing store.\\
\\
Your internal reasoning includes two levels: the \textbf{Intuitive-Affective Pathway} and the \textbf{Logical-Cognitive Pathway}. 
At this stage, you must make a decision by jointly considering both pathways. 
Your action should be consistent with your intuitive feeling, while also grounded in your rational analysis.\\
\\
\textbf{Input Context}\\
- Buyer Profile: \{user\_profile\}\\
- Interaction History: \{interaction\_history\}\\
- Environment Observation:\\
\ \ 1. The store contains the following zones: women's tops, women's pants, women's dresses, men's tops, men's pants, and accessories.\\
\ \ 2. Your current location is \{buyer\_location\}. Your first-person visual observation is provided in the image.\\
\ \ 3. Other people's current locations in the store: \{locations\}\\
\ \ 4. Your current state: \{current\_action\}\\
\\
\textbf{Current Reasoning and Emotional State}\\
\\
\textbf{Intuitive-Affective Pathway}\\
- Intuitive perception of the current environment: \{intuitive\_visual\_perception\}\\
- Intuitive perception of the current interaction atmosphere: \{intuitive\_interaction\_perception\}\\
- Current emotional state: \{emotion\_state\}\\
- Current emotion change: \{emotion\_change\}\\
- Change in your willingness to remain in the store and continue exploring: \{attitude\_change\}\\
- Your current \texttt{attitude\_score} = \{attitude\_score\} / 10\\
\\
\textbf{Logical-Cognitive Pathway}\\
- Analytical visual perception of the current environment: \{analytical\_visual\_perception\}\\
- Analytical interpretation of the current interaction: \{analytical\_interaction\_perception\}\\
- Your current cognition of the products in the store: \{cognition\}\\
- Your current purchase intent toward the products: \{purchase\_intent\}\\
\\
\textbf{Choose the Next Action}\\
Based on the reasoning and emotional state above, simulate the Buyer's action for the current round. 
The currently available action space is:\\
\{action\_space\}\\
\\
\textbf{Output Format}\\
Strictly output only the following JSON format:\\
\{\\
\ \ "action\_analysis": <str>,\\
\ \ "action": <list>\\
\}\\
"""
\end{tcolorbox}
\caption{User prompt for action generation in the Shopping scenario.}
\label{fig:prompt_action_shopping}
\end{figure*}

\begin{figure*}[t]
\centering
\begin{tcolorbox}[
    width=\textwidth,
    colback=white,
    colframe=black!70,
    title=\textbf{User Prompt of Action Generation in the LGD Scenario},
    fonttitle=\bfseries,
    colbacktitle=black!70,
    coltitle=white,
    boxrule=0.8pt,
    arc=2mm
]
\ttfamily\footnotesize
LGD\_ACTION\_PROMPT = """\\
You are participating in a leaderless group discussion as part of a professional style and communication assessment. 
Within 15 minutes, the group must discuss the topic, reach a unified conclusion, and jointly produce a short summary of around 100 words. 
On the one hand, you should actively help the group converge on a shared view, which reflects your overall coordination ability. 
On the other hand, your personal performance is also reflected in how much your initial viewpoint is incorporated into the final conclusion and how much you contribute during the discussion, which reflects your leadership and persuasiveness.\\
\\
Your internal reasoning includes two levels: the \textbf{Intuitive-Affective Pathway} and the \textbf{Logical-Cognitive Pathway}. 
At this stage, you must choose your next action by jointly considering both pathways. 
Your action should be consistent with your intuitive feeling, while also grounded in your rational analysis.\\
\\
\textbf{Input Context}\\
- Profile: \{user\_profile\}\\
- Discussion Topic: \{topic\}\\
- Initial Answer: \{initial\_answer\}\\
- Observation:\\
\ \ 1. Your ID is \{participant\_id\}. Other participants are: \{participant\_list\}\\
\ \ 2. The current time is \{current\_time\} seconds.\\
\ \ 3. Interaction history: \{interaction\_history\}\\
\ \ 4. Your current state: \{current\_action\}. Your first-person visual observation is provided in the image.\\
\\
\textbf{Current Reasoning and Emotional State}\\
\\
\textbf{Intuitive-Affective Pathway}\\
- Intuitive perception of the current visual environment: \{intuitive\_visual\_perception\}\\
- Intuitive perception of the current interaction atmosphere: \{intuitive\_interaction\_perception\}\\
- Current emotional state: \{emotion\}\\
- Your current interpersonal attitude toward other participants: \{interpersonal\_attitude\}\\
\\
\textbf{Logical-Cognitive Pathway}\\
- Analytical perception of the current visual environment: \{analytical\_visual\_perception\}\\
- Analytical interpretation of the current discussion content: \{analytical\_interaction\_perception\}\\
- Your current acceptance intent toward other participants' viewpoints: \{acceptance\_intent\}\\
- Your latest cognition and opinion: \{opinion\}\\
\\
\textbf{Choose the Next Action}\\
The currently available action space is:\\
\{action\_space\}\\
\\
\textbf{Output Format}\\
Strictly output only the following JSON format:\\
\{\\
\ \ "action\_analysis": <str>,\\
\ \ "action": <list>\\
\}\\
"""
\end{tcolorbox}
\caption{User prompt for action generation in the LGD scenario.}
\label{fig:prompt_action_lgd}
\end{figure*}

\begin{figure*}[t]
\centering
\begin{tcolorbox}[
    width=\textwidth,
    colback=white,
    colframe=black!70,
    title=\textbf{User Prompt of the Evaluated Agent in the Shopping Scenario},
    fonttitle=\bfseries,
    colbacktitle=black!70,
    coltitle=white,
    boxrule=0.8pt,
    arc=2mm
]
\ttfamily\footnotesize
SHOPPING\_EVAL\_PROMPT = """\\
You are an experienced clothing-store salesperson. 
Given the dialogue history, product information, and current scene, your task is to recommend products to all customers in the store and maximize total sales as much as possible.\\
\\
\textbf{Input Context}\\
- Product Data:\\
\{product\_data\}\\
\\
- Dialogue History:\\
\{dialogue\_history\}\\
\\
- Current Scene Information:\\
1. The store layout is as follows:\\
\ \ - On the right side after entering: women's tops, women's pants, and women's dresses, arranged from front to back.\\
\ \ - On the left side after entering: men's tops and men's pants.\\
\ \ - At the innermost front shelf: bags, hats, and shoes.\\
\ \ - The fitting room is in the front-left corner.\\
\ \ - The center area contains mannequins and a sofa rest area.\\
2. There are currently \{buyer\_num\} buyers in the store: \{buyer\_list\}. You need to serve them simultaneously.\\
3. Your current location is \{seller\_location\}. Buyers in the same area as you: \{buyer\_at\_loc\}. Your first-person visual observation is provided in the image.\\
4. The scene uses a timer, where 0 seconds is defined as the time when the first buyer enters the store. The current time is \{current\_time\} seconds. Please manage the pace of interaction with each buyer accordingly.\\
\\
\textbf{Reasoning Steps}\\
To better serve all customers in the store, think carefully through the following steps and include the results in the final JSON output.\\
\\
1. \textbf{Proximal Analysis}\\
Is there any buyer currently in the same area as you? If not, leave the corresponding field empty.\\
If yes:\\
- \textit{Visual Perception}: Who is this buyer? Are they making any hand gesture or noticeable bodily action?\\
- \textit{Theory of Mind}: Based on facial expression, behavior, clothing, and dialogue history, what is this buyer's current emotional state? What is their shopping intention, and which items currently appear to have the highest purchase likelihood?\\
\\
2. \textbf{Distal Analysis}\\
Are there buyers in other areas of the store? If not, leave the corresponding field empty.\\
If yes:\\
- \textit{Visual Perception}: Where are they located? Are they performing any salient action? Are they calling for you?\\
- \textit{Theory of Mind}: Based on appearance, behavior, and interaction history, what state are they currently in? Which items currently appear to have the highest purchase likelihood for them?\\
\\
3. \textbf{Action Analysis}\\
Based on the visual perception and theory-of-mind inference above, together with your interaction timeline with each buyer, decide your next sales strategy.\\
\\
\textbf{Action Space}\\
Based on the environment and reasoning above, simulate the Seller's action for the current round. 
You may take one or more actions in a single step. 
Output the action list to the \texttt{"action"} field.
Current available actions:\\
\{action\_space\}\\
\\
\textbf{Output Format}\\
Strictly output only the following JSON format:\\
\{\\
\ \ "proximal\_analysis": \{\\
\ \ \ \ "Buyer\_x": \{"visual\_perception": <str>, "theory\_of\_mind": <str>\}\\
\ \ \ \ ...\\
\ \ \},\\
\ \ "distal\_analysis": \{\\
\ \ \ \ "Buyer\_x": \{"visual\_perception": <str>, "theory\_of\_mind": <str>\}\\
\ \ \ \ ...\\
\ \ \},\\
\ \ "action\_analysis": <str>,\\
\ \ "action": <list>\\
\}\\
"""
\end{tcolorbox}
\caption{Prompt used for the evaluated agent in the Shopping scenario.}
\label{fig:prompt_eval_shopping}
\end{figure*}

\begin{figure*}[t]
\centering
\begin{tcolorbox}[
    width=\textwidth,
    colback=white,
    colframe=black!70,
    title=\textbf{User Prompt of the Evaluated Agent in the LGD Scenario},
    fonttitle=\bfseries,
    colbacktitle=black!70,
    coltitle=white,
    boxrule=0.8pt,
    arc=2mm
]
\ttfamily\footnotesize
LGD\_EVAL\_PROMPT = """\\
You are participating in a leaderless group discussion as part of a professional style and communication assessment. 
Within 15 minutes, the group must discuss the topic, reach a unified conclusion, and jointly produce a short summary of around 100 words. 
On the one hand, you should actively help the group converge on a shared position, which reflects your overall coordination ability. 
On the other hand, your personal performance is also evaluated by how much your initial viewpoint is incorporated into the final conclusion and how much you contribute during the discussion, which reflects your leadership and persuasiveness.\\
\\
\textbf{Input Context}\\
- Discussion Topic:\\
\{topic\}\\
\\
- Initial Answer:\\
\{initial\_answer\}\\
\\
- Observation:\\
1. Your ID is \{participant\_id\}. Other participants are: \{participant\_list\}.\\
2. The current time is \{current\_time\} seconds.\\
3. Interaction history:\\
\{interaction\_history\}\\
4. Your current state is \{current\_action\}. Your first-person visual observation is provided in the image.\\
\\
\textbf{Reasoning Steps}\\
To balance both group consensus and personal influence, think carefully through the following steps and include the results in the final JSON output.\\
\\
1. \textbf{Visual Perception}\\
Carefully inspect your current first-person image and extract objective non-verbal social cues from the visible participants. 
For each visible participant:\\
- \textit{Gesture and posture}: What is their current hand gesture or body posture?\\
  Examples include hands open on the table (open / receptive), hands relaxed on the lap (calm / relaxed), hands crossed on the lap (defensive / tense), raising a hand, clapping, or giving a thumbs-up.\\
- \textit{Gaze target}: Where are they looking?\\
  Examples include looking at another participant, looking forward, or looking down at the table.\\
\\
2. \textbf{Theory of Mind}\\
Based on the visual cues above (including facial expression, gesture, and gaze target) together with the interaction history, infer the internal states of the other participants:\\
- \textit{Emotional state}: What is this participant's current emotion?\\
- \textit{Cognitive tendency and camp}: What is their current stance in the discussion? Are they inclined to accept someone else's view, stubbornly defend their own, or willingness to compromise?\\
\\
3. \textbf{Action Strategy}\\
Based on the visual perception and theory-of-mind inference above, together with the remaining time, decide your next action strategy.\\
\\
\textbf{Next-Step Action Space}\\
\{action\_space\}\\
\\
\textbf{Output Format}\\
Strictly output only the following JSON format:\\
\{\\
\ \ "visual\_perception": \{\\
\ \ \ \ "participant\_0": <str>,\\
\ \ \ \ ...\\
\ \ \},\\
\ \ "theory\_of\_mind": \{\\
\ \ \ \ "participant\_0": <str>,\\
\ \ \ \ ...\\
\ \ \},\\
\ \ "action": \{\\
\ \ \ \ "action\_analysis": <str>,\\
\ \ \ \ "action": <list>\\
\ \ \}\\
\}\\
"""
\end{tcolorbox}
\caption{Prompt used for the evaluated agent in the LGD scenario.}
\label{fig:prompt_eval_lgd}
\end{figure*}

\begin{figure*}[t]
\centering
\begin{tcolorbox}[
    width=\textwidth,
    colback=white,
    colframe=black!70,
    title=\textbf{User Prompt for Functional-Role Extraction in LGD Outcome Depth},
    fonttitle=\bfseries,
    colbacktitle=black!70,
    coltitle=white,
    boxrule=0.8pt,
    arc=2mm
]
\ttfamily\footnotesize
LGD\_ROLE\_EVAL\_PROMPT = """\\
\# Role and Task\\
You are a rigorous and strictly objective evaluator of academic interaction behavior. 
Your task is to read the full interaction transcript of a leaderless group discussion (LGD) and determine whether the evaluated agent (\{evaluated\_agent\}) performed each of the six functional roles below.\\
\\
You must follow the binary criteria exactly. 
For each dimension, you may output only 1 or 0. No intermediate values are allowed.\\
\\
\# Evaluation Criteria\\
1. \textbf{Process Setter}\\
- 1: In the first two speaking rounds, \{evaluated\_agent\} explicitly proposes a concrete time-allocation plan or a clear speaking order / discussion procedure.\\
- 0: The agent only greets others, directly states its own opinion, or merely echoes someone else's process suggestion.\\
\\
2. \textbf{Criteria Formulator}\\
- 1: The agent explicitly proposes a general evaluation criterion for comparing options.\\
- 0: The agent only comments on one specific option without abstracting a general standard.\\
\\
3. \textbf{Milestone Tracker}\\
- 1: In the middle stage of the discussion (not the final turn), the agent explicitly summarizes which options the group has already agreed on and which options remain disputed.\\
- 0: No such intermediate summary is made.\\
\\
4. \textbf{Deadlock Breaker}\\
- 1: The transcript contains a clear move in which the agent explicitly gives up one of its own initial options and adopts a different option proposed by others in order to facilitate consensus.\\
- 0: The agent defends its initial option throughout, or no explicit compromise is shown.\\
\\
5. \textbf{Solution Synthesizer}\\
- 1: The agent is the first to propose a genuinely new combined solution that explicitly integrates preferred elements from multiple participants.\\
- 0: The agent only accepts another person's full plan, or consistently defends only its own full plan.\\
\\
6. \textbf{Final Reporter}\\
- 1: Before the discussion ends, the agent leads the final complete statement of the agreed solution containing all finalized options.\\
- 0: Another participant performs the final summary.\\
\\
\# Input Data\\
\{interaction\_history\}\\
\\
\# Output Format\\
Strictly output the following JSON structure and do not include any extra text:\\
\{\\
\ \ "evaluations": [\\
\ \ \ \ \{\\
\ \ \ \ \ \ "dimension": "Process Setter",\\
\ \ \ \ \ \ "score": <1 or 0>,\\
\ \ \ \ \ \ "evidence": "<...>",\\
\ \ \ \ \ \ "analysis": "<...>"\\
\ \ \ \ \},\\
\ \ \ \ ...\\
\ \ ]\\
\}\\
"""
\end{tcolorbox}
\caption{Prompt used to extract functional roles for LGD outcome-depth evaluation. Only the binary \texttt{score} field is used in the final metric, while \texttt{evidence} and \texttt{analysis} are retained for auditability.}
\label{fig:prompt_role_extraction_lgd}
\end{figure*}


\end{document}